*Article*

# AI-Driven Physics-Informed Bio-Silicon Intelligence System: Integrating Hybrid Systems, Biocomputing, Neural Networks, and Machine Learning, for Advanced Neurotechnology

**Vincent Jorgsson [1], Raghav Kumar [2], Mustaf Ahmed [3], Maxx Yung [4], Aryaman Pattnayak [5], Sri Pradhyumna Sridhar [6], Vaishnav Varma [7], Arun Ram Ponnambalam [8], Georg Weidlich [9], Dimitris Pinotsis [10]**

1. Synthetic Intelligence Labs: vincent.jorgsson@syntheticintelligencelabs.com, soul.syrupp@gmail.com
2. Synthetic Intelligence Labs; raghavk@syntheticintelligencelabs.com
3. Synthetic Intelligence Labs, University of Colorado Boulder; muah7432@colorado.edu
4. Synthetic Intelligence Labs, University of Pennsylvania; myung11@seas.upenn.edu
5. Synthetic Intelligence Labs, Shiv Nadar University; aryaman@syntheticintelligencelabs.com
6. Synthetic Intelligence Labs, University of Pittsburgh; sri@syntheticintelligencelabs.com
7. Synthetic Intelligence Labs, Shiv Nadar University; vaishnav@syntheticintelligencelabs.com
8. Synthetic Intelligence Labs, SRM Institute of Science & Technology; arun27pons@syntheticintelligencelabs.com
9. Synthetic Intelligence Labs, Zap Surgical Systems; georg@zapsurgical.com
10. Synthetic Intelligence Labs, Massachusetts Institute of Technology (MIT), City University of London; pinotsis@mit.edu

\* Correspondence: vincent.jorgsson@syntheticintelligencelabs.com

**Abstract:** We present the Bio-Silicon Intelligence System (BSIS), an innovative hybrid platform that integrates biological neural networks with silicon-based computing. The BSIS, a Physics-Informed Hybrid Hierarchical Reinforcement Learning State Machine, employs carbon nanotube-coated electrodes to interface rat brains with computational systems, enabling high-fidelity neural interfacing and bidirectional communication through self-organizing systems in both biological and silicon forms. Our system leverages both analogue and digital AI theory, incorporating concepts from computational theory, chaos theory, dynamical systems theory, physics, and quantum mechanics. Additionally, the BSIS replicates the neuronal dynamics typical of intelligent brain tissue, employing nonlinear operations underlying learning and information storage. Neural signals are read through the FreeEEG32 board and BrainFlow software, then features are extracted and mapped to game actions by tracking feature changes in continuous data. Metadata is encoded into both analogue and digital brain stimulation signals at the microvolt level using our proprietary software and hardware. The system employs a dual signaling approach for training the rat brain, incorporating a reward solution and sound as well as human-inaudible distress sounds. This paper details the design, theory, functionality, and technical specifications of the BSIS, highlighting its interdisciplinary approach and advanced technological integration.

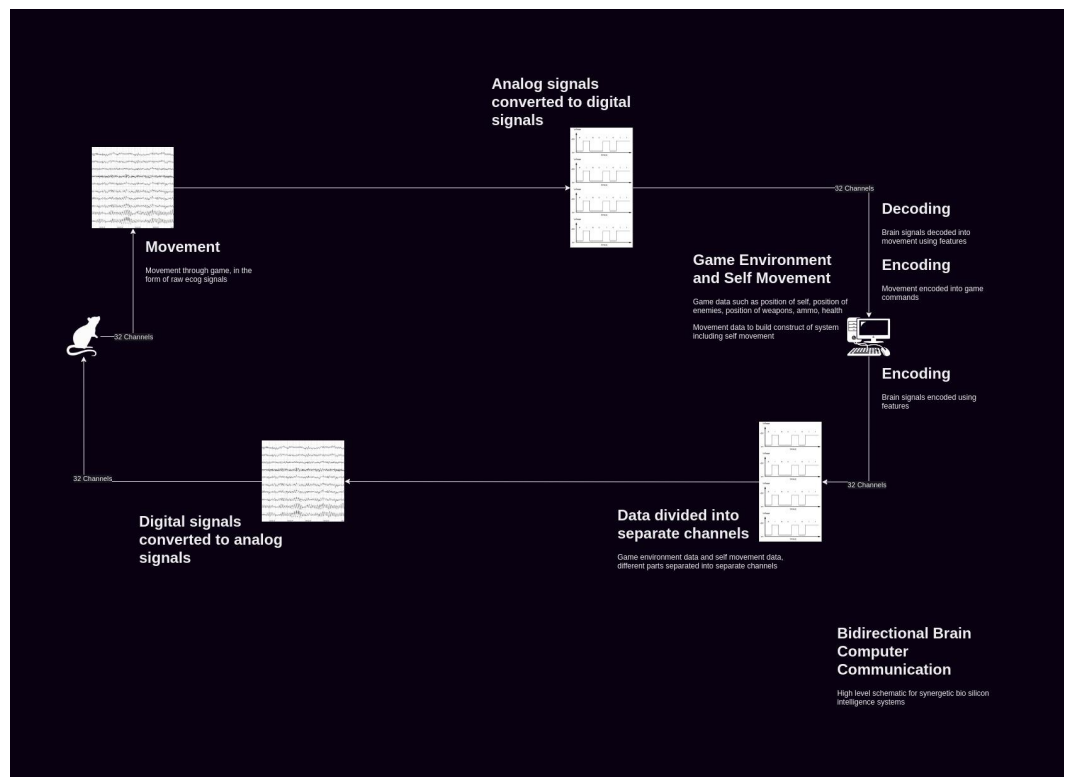

*Figure 1: Graphical Abstract*

## 1. Introduction

The integration of biological neural networks with silicon-based computing represents a significant frontier in bioengineering and neurotechnology [3,4]. Traditional computing systems, while powerful in terms of raw processing speed and storage capacity, lack the adaptability and learning capabilities inherent to biological neural networks [5,6,7,8]. For example, silicon-based systems excel at rapid calculations and data manipulation but struggle with pattern recognition and few-shot learning areas, whereas biological neural networks excel in these domains. Conversely, biological systems, despite their adaptability, face limitations in processing speed and scalability when compared to their silicon counterparts [9,10,11]. The synergy of these two paradigms—biological and silicon-based computing—presents an opportunity to harness the strengths of both to create versatile and powerful systems. This concept of bio-silicon integration forms the foundation of our research into the Bio-Silicon Intelligence System (BSIS) [12,13]. Recent advancements in brain-computer interfaces (BCIs) have opened new possibilities for direct communication between biological neurons and electronic devices [14,15]. These interfaces have the potential to revolutionize various domains, from neuroprosthetics and rehabilitation to cognitive enhancement and advanced artificial intelligence [16,17]. A critical aspect of this integration involves understanding the role of symmetry and asymmetry in nonlinear operations within neuromorphic learning mechanisms. Symmetry in neural network design can optimize learning algorithms by ensuring consistent and balanced signal processing, which enhances pattern recognition and information storage. This can lead to more efficient and effective learning processes. Conversely, asymmetry in neural networks can reveal unique properties and behaviors, such as nonlinear response dynamics and adaptive learning patterns, which are essential for replicating complex neuronal functions. In the BSIS, utilizing both symmetrical and asymmetrical principles allows for a more nuanced approach to integrating biological and silicon components.

**Innovations and System Overview**

The BSIS offers a hybrid platform that bridges the gap between biological neural networks and silicon-based computing, utilizing human cortical organoids interfaced with rat brains through carbon nanotube-coated electrodes. This integration enables high-fidelity neural interfacing and precise bidirectional communication, leveraging the natural learning and adaptability of biological systems while enhancing their capabilities with silicon-based computational power. The signal acquisition and processing framework of BSIS utilizes a Synthetic Intelligence Labs' custom MEA, FreeEEG32 board[1] and BrainFlow [2] software for accurate reading and processing of complex neural signals. The system provides comprehensive analysis and feedback using proprietary software, which includes advanced analytical tools for detailed visualization and interaction with neural data. A key innovation of BSIS is its dual signaling approach for training neural networks, employing a reward mechanism alongside human-inaudible distress sounds to guide and modify rat brain behavior. The reward mechanism reinforces desired neural patterns, while the distress sounds deter undesired activity, creating a comprehensive feedback system for shaping neural responses. This approach enhances the learning process and ensures a balanced and humane method for neural conditioning. BSIS's design incorporates robust ethical considerations, ensuring the well-being of research animals with communal housing, a carefully planned diet, and comprehensive care protocols. The versatility of BSIS supports applications in advanced artificial intelligence research, bioengineering, neuroprosthetics, cognitive enhancement, and neurotechnology, making it a powerful tool for pioneering new frontiers in these fields.

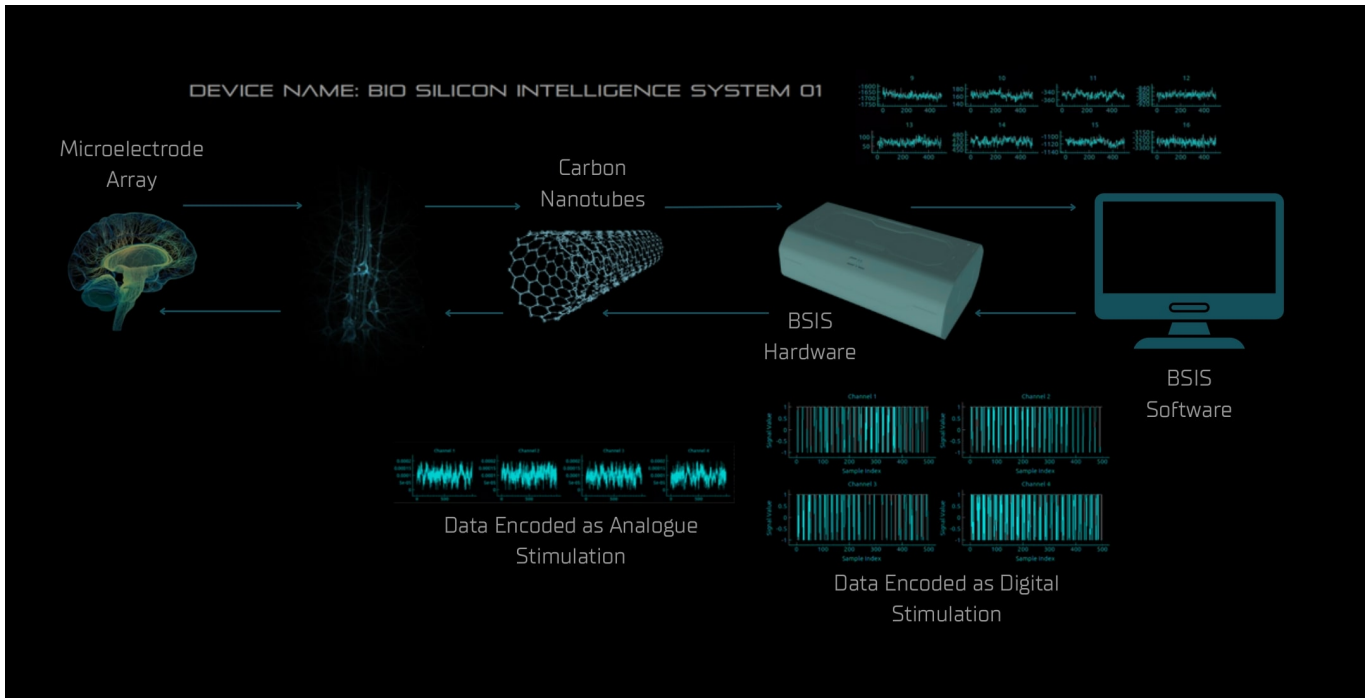

*Figure 2: System Flow of Signals*

## 2. Materials and Methods

**Physics Informed Hybrid Hierarchical Reinforcement Learning State Machine**

The Bio-Silicon Intelligence System (BSIS) functions as an advanced physics-informed Hybrid Hierarchical State Machine and analogue bio-silicon AI, effectively integrating discrete and continuous computational paradigms to manage complex neural signal processing and adaptive game control. This system synergistically combines principles from hybrid automata and hierarchical state machines, facilitating dynamic interactions between neural signals, game actions, and associated metadata. Continuous dynamics are encapsulated through an extensive array of neural features, such as peak count, spectral centroid, and Higuchi Fractal Dimension, which collectively modulate state transitions. Discrete states represent not only various game actions but also game metadata, analogue signal lambda transformations, and digital signal binary encoding. State transitions are driven by continuous neural data input, augmented by a reinforcement learning framework. The BSIS is physics-informed, meaning that its design and functionality are grounded in principles from physics, particularly in the areas of chaos theory, dynamical systems theory, and quantum mechanics. These principles inform the system's ability to process and adapt to complex neural signals in a robust and efficient manner. For instance, the system utilizes concepts from chaos theory to manage the unpredictable nature of neural signal fluctuations, while dynamical systems theory aids in understanding and modeling the continuous changes in neural states. Quantum mechanics principles are applied in the generation of neural signals, enhancing the precision and fidelity of neural interfacing.

The hierarchical architecture of BSIS organizes states across multiple levels of abstraction, efficiently managing high-level game states and nested sub-states to handle specific game mechanics, neural responses, and signal encoding processes. Reinforcement learning constitutes a pivotal component of this system, wherein the rat brain is conditioned via positive reinforcement through a nicotine-sugar-caffeine-salt water reward paired with an auditory beep, and negative reinforcement through ultrasonic distress sounds. This adaptive learning mechanism, coupled with the continuous processing of analogue signals, ensures that the system remains highly responsive and optimized for real-time neurofeedback. This physics-informed approach ensures that the BSIS can accurately and reliably interpret and respond to neural data, making it a powerful tool for real-time neurofeedback and adaptive control.

The proceeding section describes the acquisition of neural signals from rat brains using a custom multi-electrode array (MEA) coated with multi-walled carbon nanotubes (MWCNTs), ensuring high-fidelity neuron-to-electrode connections. It details the processing of signals by the FreeEEG32 board [1] and BrainFlow [2] framework. Preprocessing methods include downsampling and buffering, followed by advanced feature extraction techniques through the BSIS software such as spectral centroids, spectral edge density, Higuchi Fractal Dimension (HFD), and evolution rate. The system translates neural features into actionable commands for applications like games, offering both digital and analogue feedback options. Additionally, it outlines the stimulation hardware's role in ensuring safe neural interfacing by adjusting signal levels and minimizing electromagnetic interference.

**Signal Acquisition**

The process begins with the capture of neural signals from rat brains using a custom multi-electrode array (MEA) coated with carbon nanotubes for enhanced conductivity and biocompatibility. These electrodes provide high-fidelity signal acquisition, ensuring that neural activity is accurately detected. The captured signals are then transmitted to the FreeEEG32 board[2], which serves as the primary interface for signal acquisition. This board is equipped with 32 channels, allowing it to handle a substantial volume of neural data, ensuring high-resolution acquisition with minimal noise and distortion. Once the neural signals are captured, they are processed by the BrainFlow framework[1]. BrainFlow handles filtering and artifact removal. This ensures that the neural signals maintain their integrity and are accurately prepared for analysis and feedback. The processed signals are then fed into our proprietary software suite for comprehensive analysis. This software framework operates in conjunction with the FreeEEG32 board and BrainFlow[1,2], capturing high-resolution neural data and performing essential preprocessing tasks to maintain the integrity of the signals for further analysis.

**Multiwalled Carbon Nanotubes**

Recent research has highlighted the self-organizing properties of multi-walled carbon nanotubes (MWCNTs), which can be particularly beneficial in establishing effective neuron-to-electrode connections. MWCNTs have demonstrated the ability to spontaneously form networks that mimic natural neural pathways, promoting targeted and robust synaptic connections[18,19,20]. This self-organizing behavior helps to align the electrodes precisely with the neurons, ensuring that the neural signals are captured with high specificity and fidelity. The branching out of these connections to the right neurons is facilitated by the inherent properties of MWCNTs, which support the growth and guidance of neurites. Consequently, the system's overall performance is improved, with enhanced signal clarity and reduced noise, thereby augmenting the efficiency of neuron-electrode interfaces and contributing to more accurate and reliable neuro-computational interactions.

This integrated model for network formation and dynamics can be described as:

$$\frac{\partial P}{\partial t} = D\nabla^2 P - \beta P + \gamma \Phi P^2 - \xi \nabla \cdot (\mathbf{E} \times \mathbf{B}) + \frac{1}{\mu_0}\left(\nabla \times \mathbf{B} - \mu_0 \epsilon_0 \frac{\partial \mathbf{E}}{\partial t}\right) + \alpha \nabla \cdot (\Phi \mathbf{J}_Q)$$

$$\frac{\partial \Phi}{\partial t} = \alpha \nabla^2 \Phi + \delta P - \eta \Phi + \kappa \Phi^3 + \lambda \cos(\theta)\psi - \frac{\hbar^2}{2m}\nabla^2 \psi + V(\theta)\psi$$

$$\mathcal{H} = \int \left(\frac{1}{2}g^{ij}\partial_i \Phi \partial_j \Phi + V(\Phi) + \rho(\Phi) - \sum_{k=1}^{3}\frac{\hbar^2}{2m_k}(\nabla \psi_k)^2 + \sum_{l=1}^{n} q_l \phi_l \psi_l\right) d^3x$$

$$\nabla \cdot \mathbf{E} = \frac{\rho}{\epsilon_0} + \nabla \cdot \mathbf{P}_Q, \quad \nabla \cdot \mathbf{B} = 0, \quad \nabla \times \mathbf{E} = -\frac{\partial \mathbf{B}}{\partial t} - \frac{\partial \mathbf{P}_Q}{\partial t}$$

$$\nabla \times \mathbf{B} = \mu_0 \mathbf{J} + \mu_0 \epsilon_0 \frac{\partial \mathbf{E}}{\partial t} + \nabla \times \mathbf{M}_Q$$

$$i\hbar \frac{\partial \psi}{\partial t} = -\frac{\hbar^2}{2m}\nabla^2 \psi + V(\theta)\psi + \lambda \Phi \psi + \frac{e^2}{\epsilon_0}\sum_j |\psi_j|^2 \psi - \sum_{n=1}^{N}\frac{\partial}{\partial t}(L_n(t))$$

$$\mathbf{M} \cdot \frac{d\mathbf{x}}{dt} = \mathbf{A}\mathbf{x} + \mathbf{B}\mathbf{u} - \mathbf{C}\mathbf{y} + \nabla \cdot (\sigma \nabla \mathbf{x}) + \int_V (\mathbf{E} \cdot \nabla \rho + \mathbf{B} \cdot \nabla \mathbf{J})\, dV + \int_S (\mathbf{E} \times \mathbf{H}) \cdot d\mathbf{A}$$

Where:

- $\mu_0$ and $\epsilon_0$ are the permeability and permittivity of free space.
- $\mathbf{P}_Q$ and $\mathbf{M}_Q$ are quantum polarization and magnetization fields.
- $\psi$ is the wave function representing signal clarity.
- $\hbar$ is the reduced Planck's constant.
- $m$ is the effective mass.
- $\theta$ is the alignment angle.
- $\mathbf{M}$ is the mass matrix in the control system.
- $\mathbf{A}$, $\mathbf{B}$, and $\mathbf{C}$ are system matrices.
- $\mathbf{x}$ is the state vector.
- $\mathbf{u}$ is the input vector.
- $\mathbf{y}$ is the output vector.
- $\sigma$ is the conductivity tensor.
- $L_n(t)$ represents the inductive effects in the analogue AI system.
- $q_l$, $\phi_l$, and $\psi_l$ are charge, potential, and wave function terms in the Hamiltonian.

*Equation 1: Integrated model for network formation and dynamics*

**Real-Time Monitoring and Visualization**
   The processed and feature-extracted data are visualized in real-time through a graphical user interface (GUI). The GUI, developed using PyQt and PyQtGraph[21,22], offers an intuitive and interactive platform for monitoring neural activity. The interface displays various features such as signal amplitude, frequency bands, and other extracted metrics, allowing researchers to gain real-time insights into the neural data. To facilitate robust data transmission and integration, the BSIS software framework incorporates MQTT (Message Queuing Telemetry Transport)[23] protocols. These protocols facilitate efficient data transfer between different system components and remote servers, enabling real-time monitoring and analysis from multiple locations. The system publishes data to a central server, making it accessible for further analysis by researchers. For the BSIS system software, Figure 3 illustrates the neural signal reading display, Figure 4 the extracted features display, Figure 5 the zoomed in first section from the left of the extracted features display, Figure 6 the zoomed in second section from the left of the extracted features display, Figure 7 the third section, and Figure 8 the fourth section. Figure 9 illustrates the features to action display with averaged feature values mapped to force adjustments, and Figure 10 shows the distance to target and adjusted force over time. Figure 11 illustrates the BSIS game metadata and game action encoded analogue signals, and Figure 12 shows the game metadata and game action encoded digital signals, both digital and analogue encoded signals are sent to the BSIS stimulation hardware to stimulate the brain directly via our custom MEA.

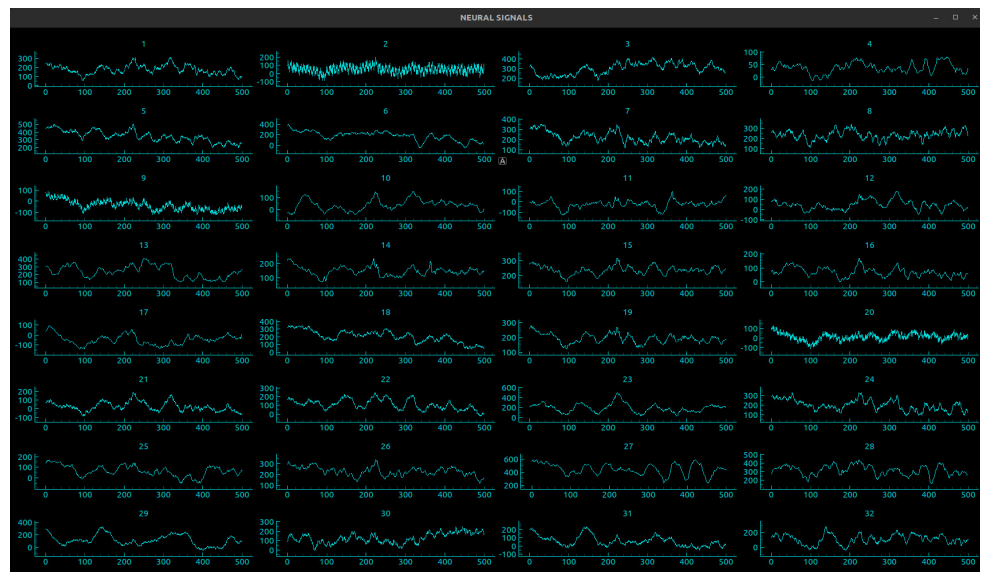

*Figure 3: Neural signals*

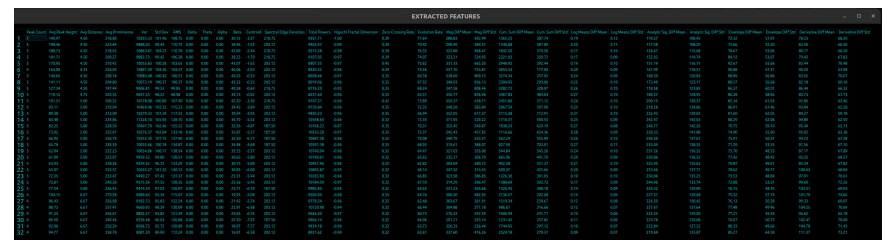

*Figure 4: Extracted Features*

| | Peak Count | Avg Peak Height | Avg Distance | Avg Prominence | Var | Std Dev | RMS | Delta | Theta | Alpha | Beta | Centroid |
|---|---|---|---|---|---|---|---|---|---|---|---|---|
| 1 | 5 | 190.97 | 4.50 | 218.80 | 10293.33 | 101.46 | 108.75 | 0.00 | 0.00 | 0.00 | 30.12 | -3.57 |
| 2 | 5 | 198.46 | 4.50 | 223.84 | 9888.03 | 99.44 | 110.72 | 0.00 | 0.00 | 0.00 | 34.46 | -1.55 |
| 3 | 5 | 188.73 | 4.50 | 218.63 | 10863.81 | 104.23 | 110.78 | 0.00 | 0.00 | 0.00 | 42.89 | -2.44 |
| 4 | 5 | 181.71 | 4.50 | 209.27 | 9883.73 | 99.42 | 106.28 | 0.00 | 0.00 | 0.00 | 39.23 | -1.70 |
| 5 | 5 | 170.95 | 4.50 | 210.42 | 10055.83 | 100.28 | 103.65 | 0.00 | 0.00 | 0.00 | 43.07 | -1.63 |
| 6 | 4 | 163.09 | 4.33 | 226.05 | 10891.59 | 104.36 | 104.57 | 0.00 | 0.00 | 0.00 | 46.06 | -2.03 |
| 7 | 5 | 149.65 | 4.50 | 208.18 | 10083.48 | 100.42 | 100.51 | 0.00 | 0.00 | 0.00 | 40.63 | -0.54 |
| 8 | 5 | 141.11 | 4.50 | 204.80 | 10073.19 | 100.37 | 100.37 | 0.00 | 0.00 | 0.00 | 43.32 | -0.33 |
| 9 | 5 | 127.34 | 4.50 | 197.44 | 9906.81 | 99.53 | 99.85 | 0.00 | 0.00 | 0.00 | 40.58 | -0.64 |
| 10 | 5 | 119.12 | 4.75 | 203.35 | 9647.25 | 98.22 | 98.98 | 0.00 | 0.00 | 0.00 | 39.13 | -0.92 |
| 11 | 5 | 101.51 | 5.00 | 200.52 | 10178.08 | 100.89 | 107.49 | 0.00 | 0.00 | 0.00 | 42.20 | -3.30 |
| 12 | 5 | 93.11 | 5.00 | 210.04 | 10469.96 | 102.32 | 115.23 | 0.00 | 0.00 | 0.00 | 39.45 | -3.89 |
| 13 | 5 | 89.38 | 5.00 | 213.09 | 10279.33 | 101.39 | 117.22 | 0.00 | 0.00 | 0.00 | 39.54 | -3.95 |
| 14 | 5 | 83.48 | 5.00 | 224.86 | 11226.10 | 105.95 | 128.10 | 0.00 | 0.00 | 0.00 | 36.79 | -5.52 |
| 15 | 5 | 82.25 | 5.00 | 228.59 | 10497.79 | 102.46 | 125.52 | 0.00 | 0.00 | 0.00 | 32.56 | -4.87 |
| 16 | 5 | 73.92 | 5.00 | 235.97 | 10576.37 | 102.84 | 133.18 | 0.00 | 0.00 | 0.00 | 35.87 | -5.57 |
| 17 | 5 | 66.95 | 5.00 | 226.79 | 10352.30 | 101.75 | 137.40 | 0.00 | 0.00 | 0.00 | 32.50 | -6.17 |
| 18 | 5 | 65.78 | 5.00 | 239.59 | 10035.66 | 100.18 | 134.87 | 0.00 | 0.00 | 0.00 | 34.48 | -4.68 |
| 19 | 5 | 62.04 | 5.00 | 222.23 | 10034.86 | 100.17 | 138.54 | 0.00 | 0.00 | 0.00 | 35.55 | -5.37 |
| 20 | 5 | 61.09 | 5.00 | 223.97 | 9959.32 | 99.80 | 138.51 | 0.00 | 0.00 | 0.00 | 30.65 | -5.80 |
| 21 | 5 | 63.93 | 5.00 | 228.36 | 9354.52 | 96.72 | 133.29 | 0.00 | 0.00 | 0.00 | 30.13 | -5.08 |
| 22 | 5 | 65.97 | 5.00 | 233.57 | 10265.27 | 101.32 | 140.12 | 0.00 | 0.00 | 0.00 | 30.83 | -6.00 |
| 23 | 5 | 72.35 | 5.00 | 233.47 | 9490.27 | 97.42 | 131.37 | 0.00 | 0.00 | 0.00 | 25.31 | -5.44 |
| 24 | 5 | 75.81 | 5.00 | 240.50 | 9515.36 | 97.55 | 128.35 | 0.00 | 0.00 | 0.00 | 25.56 | -5.45 |
| 25 | 5 | 77.54 | 5.00 | 226.43 | 9414.24 | 97.03 | 126.07 | 0.00 | 0.00 | 0.00 | 23.71 | -6.19 |
| 26 | 4 | 104.15 | 6.67 | 270.58 | 9089.64 | 95.34 | 115.67 | 0.00 | 0.00 | 0.00 | 19.95 | -5.08 |
| 27 | 4 | 96.43 | 6.67 | 256.68 | 9182.73 | 95.83 | 122.24 | 0.00 | 0.00 | 0.00 | 21.42 | -5.74 |
| 28 | 4 | 88.72 | 6.67 | 251.41 | 9660.93 | 98.29 | 130.09 | 0.00 | 0.00 | 0.00 | 22.91 | -6.68 |
| 29 | 4 | 91.23 | 6.67 | 256.61 | 8802.57 | 93.82 | 121.34 | 0.00 | 0.00 | 0.00 | 23.63 | -6.76 |
| 30 | 4 | 89.59 | 6.67 | 249.46 | 9336.48 | 96.63 | 126.88 | 0.00 | 0.00 | 0.00 | 22.50 | -7.95 |
| 31 | 4 | 92.06 | 6.67 | 250.29 | 8596.73 | 92.72 | 120.89 | 0.00 | 0.00 | 0.00 | 18.97 | -7.27 |
| 32 | 4 | 94.77 | 6.67 | 258.79 | 8081.29 | 89.90 | 112.24 | 0.00 | 0.00 | 0.00 | 16.91 | -6.58 |

*Figure 5: Extracted Features part 1*

**EXTRACTED FEATURES**

| Spectral Edge Densities | Total Powers | Higuchi Fractal Dimension | Zero Crossing Rate | Evolution Rate | Mag Diff Mean | Mag Diff Std |
|---|---|---|---|---|---|---|
| 218.75 | 9351.71 | -1.00 | 0.39 | 71.64 | 288.83 | 392.99 |
| 203.12 | 9465.97 | -0.99 | 0.39 | 70.42 | 298.49 | 384.37 |
| 218.75 | 9515.28 | -0.98 | 0.39 | 76.32 | 325.88 | 468.47 |
| 218.75 | 9107.05 | -0.97 | 0.39 | 74.07 | 323.21 | 534.95 |
| 203.12 | 8807.35 | -0.97 | 0.45 | 75.62 | 331.33 | 665.50 |
| 203.12 | 8420.32 | -0.96 | 0.29 | 74.26 | 357.76 | 753.81 |
| 203.12 | 8038.46 | -0.97 | 0.35 | 69.58 | 338.69 | 809.13 |
| 203.12 | 8019.06 | -0.96 | 0.35 | 67.52 | 348.93 | 836.13 |
| 218.75 | 8176.20 | -0.95 | 0.35 | 68.24 | 347.58 | 808.46 |
| 203.12 | 8231.60 | -0.96 | 0.35 | 63.51 | 356.77 | 836.96 |
| 218.75 | 9147.31 | -0.96 | 0.42 | 73.89 | 359.37 | 638.71 |
| 203.12 | 9759.84 | -0.96 | 0.35 | 72.35 | 348.28 | 583.84 |
| 203.12 | 9850.23 | -0.96 | 0.35 | 66.94 | 352.05 | 617.27 |
| 203.12 | 10548.64 | -0.96 | 0.32 | 73.35 | 371.95 | 529.22 |
| 187.50 | 10196.23 | -0.97 | 0.35 | 70.31 | 355.45 | 549.87 |
| 187.50 | 10533.20 | -0.97 | 0.35 | 72.37 | 345.43 | 437.85 |
| 187.50 | 10687.58 | -0.96 | 0.32 | 70.08 | 349.79 | 333.51 |
| 187.50 | 10591.38 | -0.96 | 0.35 | 68.95 | 319.61 | 388.07 |
| 203.12 | 10740.94 | -0.96 | 0.32 | 64.47 | 321.03 | 326.00 |
| 203.12 | 10740.61 | -0.96 | 0.32 | 65.65 | 332.27 | 306.70 |
| 203.12 | 10457.96 | -0.96 | 0.32 | 62.82 | 303.04 | 380.72 |
| 203.12 | 10895.87 | -0.95 | 0.32 | 68.14 | 347.32 | 314.43 |
| 203.12 | 10283.83 | -0.96 | 0.32 | 66.85 | 323.58 | 386.85 |
| 203.12 | 10184.59 | -0.97 | 0.32 | 67.10 | 314.29 | 396.47 |
| 187.50 | 9982.85 | -0.95 | 0.32 | 65.65 | 331.23 | 365.66 |
| 203.12 | 9504.94 | -0.98 | 0.35 | 64.76 | 306.50 | 482.92 |
| 203.12 | 9778.54 | -0.96 | 0.32 | 62.48 | 303.67 | 361.91 |
| 203.12 | 10120.90 | -0.94 | 0.32 | 66.44 | 304.88 | 271.18 |
| 203.12 | 9666.50 | -0.97 | 0.32 | 60.73 | 270.33 | 347.59 |
| 187.50 | 9966.14 | -0.96 | 0.32 | 66.06 | 291.21 | 293.14 |
| 203.12 | 9434.18 | -0.96 | 0.32 | 63.73 | 326.33 | 326.46 |
| 203.12 | 8921.62 | -0.99 | 0.32 | 63.61 | 337.60 | 416.26 |

*Figure 6: Extracted Features part 2*

| Cum. Sum Diff Mean | Cum. Sum Diff Std | Log Means Diff Mean | Log Means Diff Std | Analytic Sig. Diff Mean | Analytic Sig. Diff Std |
|---|---|---|---|---|---|
| 1365.25 | 387.74 | 0.19 | 0.12 | 119.27 | 108.45 |
| 1340.88 | 387.89 | 0.20 | 0.11 | 117.58 | 108.20 |
| 1842.50 | 379.56 | 0.17 | 0.10 | 126.67 | 115.68 |
| 2321.03 | 359.72 | 0.17 | 0.09 | 132.03 | 114.74 |
| 2948.92 | 392.44 | 0.19 | 0.10 | 151.19 | 116.71 |
| 3055.53 | 177.00 | 0.20 | 0.09 | 161.99 | 118.51 |
| 3274.34 | 257.92 | 0.24 | 0.09 | 168.39 | 120.83 |
| 3304.95 | 293.06 | 0.25 | 0.10 | 173.49 | 123.17 |
| 3282.72 | 328.97 | 0.26 | 0.10 | 178.58 | 123.83 |
| 3467.83 | 484.04 | 0.27 | 0.10 | 189.27 | 128.45 |
| 2451.86 | 571.12 | 0.24 | 0.10 | 200.15 | 130.37 |
| 2067.54 | 597.90 | 0.25 | 0.10 | 218.28 | 134.86 |
| 2115.28 | 712.91 | 0.31 | 0.10 | 232.43 | 139.02 |
| 1710.27 | 590.55 | 0.25 | 0.09 | 242.47 | 142.65 |
| 1838.24 | 629.19 | 0.27 | 0.08 | 246.77 | 142.26 |
| 1116.66 | 654.36 | 0.28 | 0.09 | 250.33 | 141.88 |
| 562.24 | 555.99 | 0.26 | 0.10 | 246.50 | 137.61 |
| 827.90 | 703.91 | 0.27 | 0.11 | 253.04 | 138.35 |
| 544.84 | 545.26 | 0.24 | 0.10 | 251.26 | 136.52 |
| 665.96 | 441.70 | 0.20 | 0.09 | 250.86 | 136.52 |
| 992.28 | 551.37 | 0.21 | 0.10 | 252.95 | 135.63 |
| 699.37 | 425.66 | 0.20 | 0.09 | 253.66 | 137.71 |
| 1226.30 | 391.05 | 0.19 | 0.10 | 250.06 | 135.25 |
| 1388.69 | 392.73 | 0.17 | 0.09 | 244.96 | 133.74 |
| 1326.45 | 308.18 | 0.14 | 0.09 | 235.32 | 129.99 |
| 2136.57 | 292.68 | 0.14 | 0.09 | 227.51 | 129.68 |
| 1519.34 | 259.67 | 0.12 | 0.08 | 226.50 | 130.42 |
| 986.67 | 316.66 | 0.12 | 0.07 | 227.67 | 131.64 |
| 1606.44 | 241.71 | 0.10 | 0.06 | 225.55 | 129.30 |
| 1231.45 | 257.82 | 0.11 | 0.06 | 224.78 | 129.08 |
| 1744.95 | 297.12 | 0.10 | 0.07 | 222.84 | 127.52 |
| 2524.18 | 279.31 | 0.09 | 0.07 | 219.69 | 125.87 |

*Figure 7: Extracted Features part 3*

| Envelope Diff Mean | Envelope Diff Std | Derivative Diff Mean | Derivative Diff Std |
|---|---|---|---|
| 72.32 | 51.97 | 78.23 | 68.20 |
| 75.66 | 52.20 | 83.50 | 66.50 |
| 78.67 | 53.06 | 80.17 | 66.59 |
| 84.12 | 53.67 | 79.43 | 67.83 |
| 92.67 | 55.68 | 83.44 | 70.48 |
| 90.88 | 57.31 | 90.20 | 63.98 |
| 88.89 | 56.86 | 83.92 | 70.07 |
| 88.57 | 56.68 | 82.18 | 69.18 |
| 86.37 | 60.01 | 86.44 | 66.32 |
| 86.28 | 58.66 | 85.73 | 61.73 |
| 85.34 | 61.93 | 92.86 | 63.46 |
| 86.01 | 61.46 | 93.84 | 62.28 |
| 81.60 | 62.05 | 84.27 | 59.18 |
| 88.29 | 62.08 | 99.89 | 62.59 |
| 79.75 | 59.10 | 95.49 | 62.13 |
| 74.90 | 55.90 | 95.02 | 63.38 |
| 75.91 | 50.37 | 94.23 | 67.28 |
| 71.29 | 53.55 | 91.56 | 67.10 |
| 73.70 | 48.33 | 87.17 | 67.89 |
| 77.42 | 48.45 | 93.25 | 68.57 |
| 70.87 | 49.61 | 85.34 | 67.82 |
| 78.62 | 49.77 | 100.63 | 68.84 |
| 73.53 | 48.84 | 97.91 | 70.61 |
| 72.88 | 53.51 | 99.60 | 72.26 |
| 76.15 | 48.76 | 103.51 | 69.03 |
| 79.32 | 57.74 | 101.79 | 74.66 |
| 76.13 | 50.58 | 99.33 | 69.07 |
| 77.48 | 49.46 | 104.25 | 70.69 |
| 77.21 | 45.43 | 96.62 | 65.18 |
| 76.67 | 50.73 | 102.47 | 70.68 |
| 80.33 | 49.66 | 104.78 | 71.43 |
| 85.27 | 64.58 | 111.37 | 73.21 |

*Figure 8: Extracted Features part 4*

| | Feature | Value | Action |
|---|---|---|---|
| 1 | variance | 15867.8086 | adjust_force (1) |
| 2 | std_dev | 125.9675 | adjust_force (1) |
| 3 | rms | 133.1937 | adjust_force (1) |
| 4 | peak_count | 4.0000 | adjust_force (0) |
| 5 | average_peak_height | 217.7075 | adjust_force (-1) |
| 6 | average_distance | 8.0000 | adjust_force (0) |
| 7 | average_prominence | 266.6488 | adjust_force (1) |
| 8 | beta_band_power | 38.4582 | adjust_force (1) |
| 9 | centroids | -5.1049 | adjust_force (1) |
| 10 | spectral_edge_densities | 218.7500 | adjust_force (-1) |
| 11 | higuchi_fractal_dimension | -0.9679 | adjust_force (1) |
| 12 | zero_crossing_rate | 0.4194 | adjust_force (-1) |
| 13 | magnitudes_diff_mean | 322.9809 | adjust_force (-1) |
| 14 | magnitudes_diff_std | 458.0879 | adjust_force (-1) |
| 15 | cumulative_sums_diff_me... | 817.7864 | adjust_force (-1) |
| 16 | cumulative_sums_diff_std | 749.4072 | adjust_force (1) |
| 17 | log_means_diff_mean | 0.2733 | adjust_force (-1) |
| 18 | log_means_diff_std | 0.1080 | adjust_force (1) |
| 19 | analytic_signals_diff_mean | 93.8639 | adjust_force (1) |
| 20 | analytic_signals_diff_std | 90.4118 | adjust_force (1) |
| 21 | envelopes_diff_mean | 68.1675 | adjust_force (1) |
| 22 | envelopes_diff_std | 72.7703 | adjust_force (1) |
| 23 | derivatives_diff_mean | 65.0353 | adjust_force (1) |
| 24 | derivatives_diff_std | 60.6492 | adjust_force (-1) |

Total Force Adjustment: -6.88

*Figure 9: Force adjustments from mapped features to game actions*

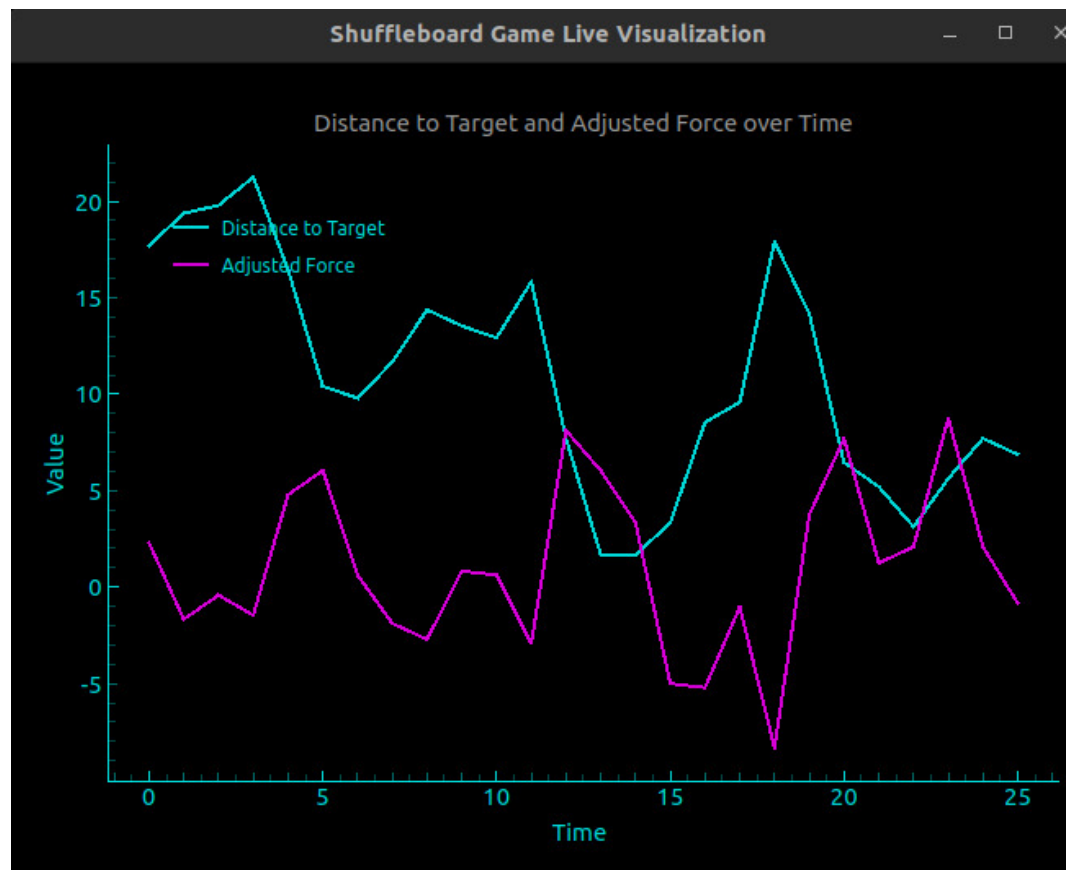

*Figure 10: Distance to target and adjusted force over time*

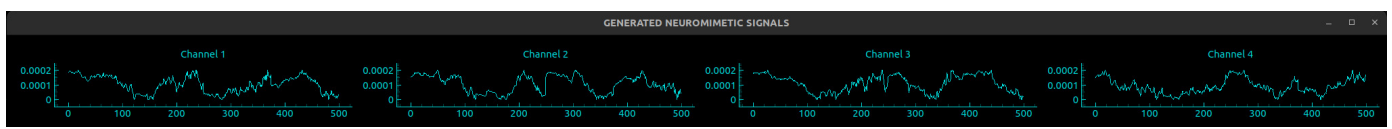

*Figure 11: Game metadata and game action encoded analogue signals for 250ms at 48000hz upsampled from 500hz*

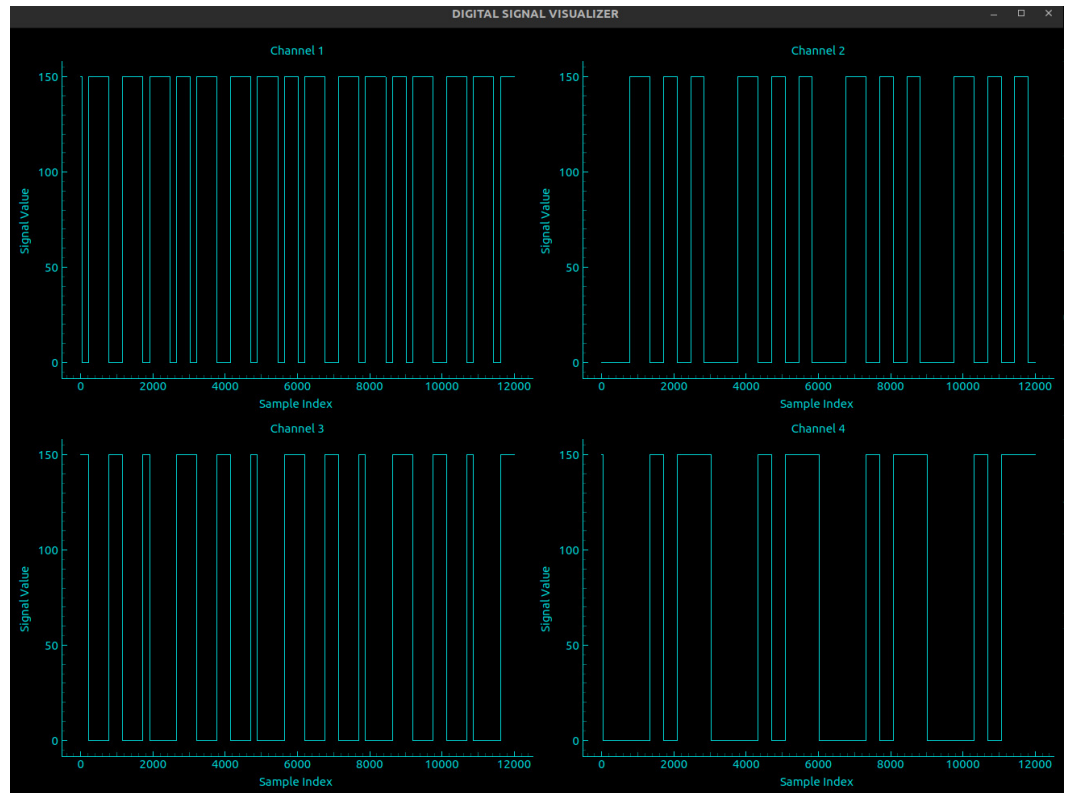

*Figure 12: Game metadata and game action encoded digital signals for 250ms upsampled to 48000hz*

**Preprocessing and Buffering**
  The preprocessing stage of BSIS begins immediately after neural signal capture. The system downsamples the capture signals to a manageable rate for processing. Downsampling is performed using resampling techniques to preserve the essential characteristics of the data while reducing the computational load on the system. The processed data is then stored in a buffer, which maintains a continuous stream of recent neural activity. This buffered data serves as the input for further analysis.

**Theory, Feature Extraction, and Calculation**
**Spectral Centroids**
  The concept of spectral centroids is deeply rooted in the mathematical framework of signal processing and Fourier analysis. The spectral centroid represents the "center of mass" of a signal's power spectrum, which can be derived by transforming the signal from the time domain to the frequency domain using the Fast Fourier Transform (FFT) [24,25]. The FFT is based on the Discrete Fourier Transform (DFT), which converts a finite sequence of equally spaced samples of a function into a same-length sequence of equally spaced samples of the discrete-time Fourier transform (DTFT), which is a complex-valued function of frequency. The FFT's computational efficiency (O(N log N) vs O(N^2) for direct DFT calculation) makes it practical for real-time signal processing applications, such as those in BSIS.

The theoretical foundation of the FFT relies on the orthogonality of sine and cosine functions, which form an orthogonal basis for the space of periodic functions. This orthogonality ensures that each frequency component in the signal can be independently analyzed, providing a precise decomposition of the signal into its constituent frequencies. This decomposition is critical for accurately determining the spectral centroid, as it allows for the clear identification and quantification of the power associated with each frequency component.

The spectral centroid $C$ is calculated using the following formula:

$$C = \frac{\sum_{i=1}^{N} f_i \cdot m_i}{\sum_{i=1}^{N} m_i}$$

Where:

- $C$ is the spectral centroid
- $f_i$ are the discrete frequencies obtained from the FFT
- $m_i$ are the magnitudes of the corresponding frequency components
- $N$ is the total number of frequency components

This weighted average computes a summary of where the bulk of the signal's energy is concentrated, with the weights being the magnitudes of each frequency component.

*Equation 2: Spectral Centroids*

In theoretical physics, particularly in the study of wave phenomena, the spectral centroid can be interpreted as the average energy state of a system. For instance, in quantum mechanics, the energy levels of a quantum system are analogous to the frequency components of a signal. The spectral centroid, therefore, represents an average energy state, providing insights into the system's overall energy distribution.

This analogy is powerful in understanding the behavior of complex systems, where energy distribution and transitions between states are critical aspects [26].

The average energy state $\langle E \rangle$ in theoretical physics, particularly in the study of wave phenomena, can be interpreted using the following formula:

$$\langle E \rangle = \frac{\int_0^\infty E(\omega)\rho(\omega)\,d\omega}{\int_0^\infty \rho(\omega)\,d\omega} = \frac{\int_0^\infty \hbar\omega\,|\psi(\omega)|^2\,d\omega}{\int_0^\infty |\psi(\omega)|^2\,d\omega}$$

Where:

- $\langle E \rangle$ is the spectral centroid, representing the average energy state of the system.
- $E(\omega)$ is the energy at angular frequency $\omega$.
- $\rho(\omega)$ is the energy density function.
- $\hbar$ is the reduced Planck's constant.
- $\omega$ is the angular frequency.
- $\psi(\omega)$ is the wave function in the frequency domain.

*Equation 3: Energy State*

The spectral centroid $\langle\omega\rangle$ also finds applications in the analysis of non-linear systems and chaotic signals. In such systems, the power distribution across frequencies can reveal underlying patterns and structures that are not immediately apparent in the time domain. By examining the spectral centroid, researchers can gain insights into the dominant frequencies and their contributions to the overall system behavior, facilitating a deeper understanding of the dynamics at play [27,28,29]. For example, computing the spectral centroid can assist in detecting the onset of chaos in dynamical systems and characterize the overall frequency content of chaotic attractors.

$$\langle\omega\rangle = \frac{\int_0^\infty \omega |\mathcal{F}\{\phi(t)\}(\omega)|^2 \, d\omega + \int_{\mathbb{R}^3} \mathbf{k} \cdot \nabla |\psi(\mathbf{r},t)|^2 \, d^3r}{\int_0^\infty |\mathcal{F}\{\phi(t)\}(\omega)|^2 \, d\omega + \int_{\mathbb{R}^3} |\psi(\mathbf{r},t)|^2 \, d^3r}$$

Where:

- $\langle\omega\rangle$ is the spectral centroid, representing the average frequency of the system.
- $\omega$ is the angular frequency.
- $\phi(t)$ is the time-domain signal representing the system's state.
- $\mathcal{F}\{\phi(t)\}(\omega)$ is the Fourier transform of $\phi(t)$, mapping the time-domain signal to the frequency domain.
- $\psi(\mathbf{r},t)$ is the wave function representing the system's state in both space and time.
- $\mathbf{k}$ is the wave vector.
- $\mathbf{r}$ is the position vector in three-dimensional space.

*Equation 4: Spectral Centroid for Frequency Analysis in Non-Linear and Chaotic Systems*

In the context of neural signals, spectral centroids can provide insights into the dominant frequency ranges of neural activity. Shifts in the spectral centroid can indicate changes in cognitive states or the engagement of different neural networks. For example, a shift towards higher frequencies might indicate increased cognitive load or alertness while a shift towards lower frequencies could suggest relaxation or drowsiness. However, while sufficient for the BSIS system, spectral centroids have several limitations, including high sensitivity to signal noise.

**Spectral Edge Density**

Spectral edge density is a measure that captures the distribution of power within a signal's spectrum up to a specified threshold, typically 95% of the total power[30]. The calculation of spectral edge density involves transforming the signal into the frequency domain using the FFT, which reveals the signal's power spectrum. The cumulative power distribution is then computed by summing the power contributions of individual frequency components, starting from the lowest frequency.

Theoretical exploration of spectral edge density involves the use of cumulative distribution functions (CDFs) and quantile analysis[31,32]. The cumulative power distribution can be seen as analogous to a CDF, representing the accumulation of power across the frequency spectrum [33,34].

The spectral edge is identified as the frequency $f_{\text{edge}}$ below which a specified percentage of the total power is contained. Mathematically, this can be expressed as:

$$f_{\text{edge}} = f \quad \text{such that} \quad \frac{\sum_{i=1}^{k} m_i}{\sum_{i=1}^{N} m_i} \geq \frac{P}{100}$$

Where:

- $f_{\text{edge}}$ is the spectral edge frequency.
- $f_i$ are the discrete frequencies obtained from the FFT.
- $m_i$ are the magnitudes of the corresponding frequency components.
- $N$ is the total number of frequency components.
- $k$ is the index where the cumulative sum of magnitudes reaches or exceeds the specified percentage $P$ of the total power.

This measure provides insights into the concentration and dispersion of signal power, highlighting the frequencies that contain the majority of the signal's energy.

*Equation 5: Spectral Edge*

This measure provides insights into the concentration and dispersion of signal power, highlighting the frequencies that contain the majority of the signal's energy. In theoretical physics, spectral edge density can be linked to thermodynamic principles and statistical mechanics. In these fields, the distribution of particles or energy states within a system can be described by cumulative distributions, with the spectral edge corresponding to a specific quantile of this distribution. This analogy provides a framework for understanding how energy is distributed within a system and the thresholds that characterize significant energy concentrations[35].

In the context of signal processing and theoretical neuroscience, spectral edge density is crucial for understanding how signal power is spread across frequencies. This measure is particularly valuable in distinguishing between different types of signals based on their power distribution characteristics. For example, in brain signal analysis, different cognitive states or neural activities can exhibit distinct spectral edge densities, reflecting the underlying neural dynamics and connectivity [36,37]. Furthermore, spectral edge density can be related to concepts in wave mechanics and acoustics. In these fields, the distribution of energy across different frequency components determines the behavior and properties of waves. The spectral edge, therefore, provides a critical parameter for characterizing wave phenomena and their interactions with the environment. By understanding the spectral edge, researchers can gain insights into the propagation and dispersion of waves, as well as the concentration of energy within specific frequency bands [38].

This measure provides insights into the concentration and dispersion of signal power, highlighting the frequencies that contain the majority of the signal's energy. In theoretical physics, spectral edge density can be linked to thermodynamic principles and statistical mechanics. In these fields, the distribution of particles or energy states within a system can be described by cumulative distributions, with the spectral edge corresponding to a specific quantile of this distribution. This analogy provides a framework for understanding how energy is distributed within a system and the thresholds that characterize significant energy concentrations[35].

**Integration of Theoretical Concepts**

The theoretical underpinnings of spectral centroids and spectral edge density highlight their significance in the broader context of signal analysis. These features encapsulate fundamental principles from Fourier analysis, energy distribution, and statistical mechanics, providing deep insights into the structure and dynamics of signals. By capturing the essence of power and frequency distribution, these measures bridge the gap between mathematical theory and practical applications, offering powerful tools for the analysis and interpretation of complex signals in various scientific fields.

The theoretical depth of spectral centroids and spectral edge density extends beyond simple calculations, delving into the core principles of signal processing, theoretical physics, and energy distribution. These measures offer profound insights into the behavior and characteristics of signals, providing a robust framework for understanding and analyzing complex systems. Through their applications in fields like neuroscience, acoustics, and wave mechanics, spectral centroids and spectral edge density demonstrate their versatility and importance in advancing our understanding of dynamic systems and their underlying processes [39].

**Higuchi Fractal Dimension**

The Higuchi Fractal Dimension (HFD)[40] provides a quantitative measure of the complexity of a signal, with higher values indicating greater complexity and irregularity. The theoretical foundation of HFD lies in chaos theory and the study of dynamical systems. In these fields, fractal dimensions are used to characterize the behavior of chaotic systems, which are sensitive to initial conditions and exhibit complex, aperiodic behavior. The HFD measures the self-similarity and intricacy of a signal, capturing the essence of this chaotic behavior. This allows for a more nuanced understanding of systems that do not exhibit regular periodicity but are instead governed by underlying non-linear dynamics. Such insights are invaluable in fields where understanding the unpredictable behavior of a system can lead to better models and predictions.

In the realm of theoretical physics, the concept of fractals extends to the study of natural phenomena and structures that exhibit fractal properties. For instance, the distribution of galaxies in the universe, the structure of turbulent flows, and the morphology of biological systems all exhibit fractal characteristics. The HFD, therefore, serves as a bridge between signal processing and the analysis of complex, natural systems, providing a tool for understanding the underlying mechanisms that drive these phenomena. By quantifying the self-similar and scale-invariant properties of such systems, the HFD aids in uncovering the fundamental principles that govern their formation and evolution. This connection between fractal geometry and natural phenomena underscores the universality of fractal dimensions in describing complex structures [41,42,43,44].

In neuroscience, the application of HFD is particularly significant for analyzing neural signals, which often exhibit complex, non-linear dynamics. Neural activity, characterized by oscillations and transient events, can be effectively studied using fractal dimensions. The HFD provides insights into the complexity of neural signals, reflecting the underlying neural connectivity and functional organization. Higher HFD values in neural signals may indicate more intricate neural processing and higher levels of neural interactivity. This is particularly useful for distinguishing between different states of brain activity, such as rest versus task engagement, and can also aid in identifying pathological conditions characterized by abnormal neural dynamics. The ability of the HFD to capture the intricate patterns in neural signals makes it a valuable tool in both research and clinical settings.

Moreover, the HFD can be linked to the concept of self-organized criticality (SOC), a property of dynamical systems that naturally evolve to a critical state where minor events can lead to significant, system-wide changes. SOC is observed in various natural systems, including neural networks, where it manifests as spontaneous activity that exhibits scale-invariant behavior. The fractal dimension, including the HFD, captures this scale-invariant property, providing a measure of the system's propensity for criticality and complexity. This theoretical framework explains how neural systems can maintain a delicate balance between stability and flexibility, enabling both robust information processing and adaptability. Understanding SOC through the lens of HFD enhances our comprehension of how complex neural dynamics emerge from simple, local interactions [45,46,47].

The practical application of HFD in our system involves several critical steps, each grounded in the theoretical principles discussed. First, the neural signal is segmented into smaller sections to facilitate multi-scale analysis. The length of the signal within these segments is then computed across different scales, capturing the self-similarity of the signal.

This process leverages the recursive nature of fractals, where patterns repeat at different scales, to quantify the complexity of the signal. Specifically, the calculation involves:

- Segmenting the Signal: The signal is divided into segments to facilitate analysis at various scales. Each segment's length is calculated by measuring the cumulative distances between points at different intervals, capturing the fractal nature of the signal.

- Calculating Length at Various Scales: For each segment, the length $L_k$ is computed for different values of $k$, where $k$ represents the scale. The length at scale $k$ is given by:

$$L_k(m) = \frac{1}{k} \left( \sum_{i=1}^{\lfloor \frac{N-m}{k} \rfloor} |x(m + i \cdot k) - x(m + (i-1) \cdot k)| \right) \cdot \frac{N-1}{\lfloor \frac{N-m}{k} \rfloor \cdot k}$$

Where $x(t)$ is the signal, $N$ is the total length of the signal, and $m$ varies from 1 to $k$.

- Log-Log Plot Construction: The lengths $L_k$ are then averaged over different starting points to account for variability, and a log-log plot of $L_k$ versus $k$ is constructed. This plot provides a visual representation of how the signal's complexity scales with $k$.

- Estimating Fractal Dimension: The slope of the log-log plot, obtained through linear regression, provides the Higuchi Fractal Dimension $D$. The slope is calculated using:

$$D = \frac{\log L(k)}{\log \left( \frac{1}{k} \right)}$$

This fractal dimension quantifies the complexity of the signal, with higher values indicating greater irregularity and self-similarity.

*Equation 6: Higuchi Fractal Dimension*

The Higuchi Fractal Dimension is a powerful tool for quantifying the complexity of signals, rooted in the principles of fractal geometry, chaos theory, and dynamical systems. By analyzing the self-similarity of a signal across different scales, the HFD provides deep insights into the structure and behavior of complex systems, bridging the gap between theoretical mathematics, physics, and practical applications in signal processing and neuroscience. Through its ability to capture the intricacy and irregularity of signals, the HFD stands as a testament to the profound interconnectedness of mathematical theory and the natural world. This interconnectedness is further exemplified by the practical applications of HFD in our system, where it plays a crucial role in feature extraction and the analysis of neural signals, providing valuable insights into the dynamics of brain activity.

**Evolution Rate**

The evolution rate captures the dynamic behavior of a signal by quantifying its rate of change over time. This concept is deeply rooted in several theoretical frameworks, including complex analysis, signal processing, theoretical physics, and neuroscience. The process involves the Hilbert transform[48,49], which converts a real-valued signal into its analytic signal, providing both amplitude and phase information. This transformation allows for a detailed examination of the signal's temporal dynamics.

**Hilbert Transform and Analytic Signal**

The Hilbert transform is an integral operator that shifts the phase of a signal by 90 degrees, producing an imaginary component from a real-valued signal. This results in the creation of the analytic signal, z(t)=x(t)+ix^(t), where x^(t) is the Hilbert transform of x(t). The Hilbert transform is defined by:

Mathematically, the calculation involves the following steps:

1. **Hilbert Transform**: The Hilbert transform $\mathcal{H}\{x(t)\}$ of a real-valued signal $x(t)$ converts it into its analytic signal $z(t)$:

$$z(t) = x(t) + j\mathcal{H}\{x(t)\}$$

where $j$ is the imaginary unit.

2. **Envelope Calculation**: The envelope $A(t)$ of the analytic signal $z(t)$ is given by:

$$A(t) = |z(t)| = \sqrt{x(t)^2 + \mathcal{H}\{x(t)\}^2}$$

3. **Derivative of Envelope**: The rate of change of the envelope is calculated using its discrete derivative:

$$\frac{dA(t)}{dt} \approx A(t_{i+1}) - A(t_i)$$

4. **Evolution Rate**: The evolution rate $R$ is then computed as the mean of the absolute values of these discrete derivatives:

$$R = \frac{1}{N-1} \sum_{i=1}^{N-1} |A(t_{i+1}) - A(t_i)|$$

Where: - $N$ is the total number of samples in the signal. - $A(t)$ is the envelope of the analytic signal.

This measure provides insights into the dynamic behavior of the signal by quantifying the rate of change of its amplitude over time.

*Equation 7: Hilbert Transform-Based Method for Signal Evolution Rate Calculation*

This integral is taken as the Cauchy principal value, which ensures convergence and proper handling of singularities. The analytic signal is a complex-valued function that facilitates the separation of the instantaneous amplitude and phase. The envelope represents the instantaneous amplitude of the signal, capturing its amplitude variations independently of phase information.

In the context of signal processing, the instantaneous amplitude (envelope) is critical for understanding the energy distribution of the signal. The energy of a signal is proportional to the square of its amplitude, hence variations in the envelope directly reflect changes in the signal's energy. Theoretical foundations in energy modulation and complex signal analysis emphasize the importance of the envelope in representing the signal's power dynamics. The derivative of the envelope with respect to time, $\frac{dA(t)}{dt}$, quantifies the rate at which the signal's amplitude evolves. This rate of change provides insights into the temporal dynamics of the signal, highlighting periods of rapid fluctuation and stability. In theoretical terms, the derivative captures the instantaneous rate of energy transfer within the signal, a concept that is fundamental in the study of dynamic systems [50,51,52].

The evolution rate's theoretical significance extends to the field of dynamical systems and chaos theory. Dynamical systems describe how the state of a system evolves over time according to a set of deterministic rules. These systems can exhibit a range of behaviors from stable periodic oscillations to chaotic dynamics, where small changes in initial conditions can lead to vastly different outcomes. The evolution rate, by quantifying the amplitude changes in a signal, provides a measure of the system's dynamic behavior. In chaotic systems, the sensitivity to initial conditions and the presence of strange attractors result in complex, aperiodic behavior. The rate of change of the envelope can capture these chaotic dynamics, revealing the underlying complexity of the signal. The Hilbert transform, by providing a complex representation of the signal, enables the analysis of these intricate behaviors in the complex plane [53,54].

$$A(t) = |\mathcal{H}\{x(t)\}| = |x(t) + i\mathcal{H}\{x(t)\}|$$

$$\frac{dA(t)}{dt} = \frac{d}{dt}|x(t) + i\mathcal{H}\{x(t)\}| = \frac{\Re\{x(t) + i\mathcal{H}\{x(t)\}\} \cdot \frac{dx(t)}{dt} + \Im\{x(t) + i\mathcal{H}\{x(t)\}\} \cdot \frac{d\mathcal{H}\{x(t)\}}{dt}}{|x(t) + i\mathcal{H}\{x(t)\}|}$$

Where:

- $A(t)$ is the envelope of the signal $x(t)$.
- $\mathcal{H}\{x(t)\}$ is the Hilbert transform of $x(t)$, providing the imaginary part of the analytic signal.
- $\frac{dA(t)}{dt}$ represents the evolution rate, or the rate of change of the envelope with respect to time.
- $\Re\{\cdot\}$ and $\Im\{\cdot\}$ denote the real and imaginary parts, respectively.
- $\frac{dx(t)}{dt}$ and $\frac{d\mathcal{H}\{x(t)\}}{dt}$ are the time derivatives of $x(t)$ and its Hilbert transform.

*Equation 8: Derivative of Envelope Using Hilbert Transform for Signal Evolution Analysis*

This formulation provides a comprehensive view of the signal's amplitude dynamics, capturing both instantaneous amplitude and its rate of change. The Hilbert transform is crucial in this analysis, offering a means to understand the signal's behavior in the complex plane. By examining the envelope and its derivative, researchers can gain deep insights into the energy distribution and dynamic behavior of non-linear and chaotic systems. In theoretical neuroscience, the evolution rate is crucial for analyzing the dynamic behavior of neural signals. Neural activity is inherently dynamic, characterized by oscillations, spikes, and transient events. The evolution rate offers a quantitative measure of these dynamics, reflecting the underlying neural processes. Theoretical models of neural activity often involve differential equations and dynamical systems theory to describe the interactions within neural networks. The evolution rate can provide insights into neural connectivity and functional organization. For example, rapid changes in the evolution rate may indicate heightened neural activity associated with sensory processing, cognitive tasks, or pathological conditions like epilepsy. Theoretical neuroscience posits that neural signals are manifestations of complex network interactions, and the evolution rate captures these interactions' temporal characteristics [55].

Self-organized criticality (SOC) is a fundamental property of dynamical systems, including neural networks, that naturally evolve towards a critical state where minor events can trigger significant, system-wide changes. This phenomenon, first described by Bak, Tang, and Wiesenfeld in 1987, has since been observed across a wide range of natural systems, from earthquakes and forest fires to economic markets and, notably, neural networks. In neural systems, SOC manifests as a spontaneous activity that exhibits scale-invariant behavior, where patterns are self-similar across different temporal and spatial scales. The evolution rate captures these scale-invariant properties by measuring changes in the signal's amplitude. In SOC systems, the amplitude fluctuations reflect the system's propensity for criticality and complexity. The theoretical framework of SOC provides a basis for understanding how neural systems maintain a balance between stability and flexibility, enabling both robust information processing and adaptability [56,57,58].

In our system, the evolution rate is calculated by first obtaining the analytic signal using the Hilbert transform. The envelope of the signal is then extracted, and the rate of change of this envelope is computed. The mean of the absolute values of these changes provides a single measure of the signal's dynamic behavior. This practical implementation is grounded in the theoretical principles discussed, ensuring that the measure accurately captures the temporal dynamics of the signal. The evolution rate is particularly valuable for distinguishing between different types of neural activity, identifying transient events, and understanding the overall temporal dynamics of neural signals. In neural systems exhibiting SOC, amplitude fluctuations are not random but follow specific statistical distributions, often characterized by power laws. These power law distributions reflect the system's balance between order and chaos. By providing a detailed view of the signal's behavior, the evolution rate enhances our ability to analyze and interpret complex neural data. The evolution rate is a powerful feature rooted in deep theoretical concepts from complex analysis, signal processing, theoretical physics, and neuroscience. By capturing the rate of change in signal amplitude, it provides profound insights into the dynamic behavior of neural signals. This feature bridges the gap between theoretical understanding and practical application, offering a robust tool for analyzing the temporal dynamics of complex systems.

$$R_e = \frac{1}{T} \int_0^T \left| \frac{d}{dt} \left( \sqrt{x(t)^2 + \mathcal{H}\{x(t)\}^2} \right) \right| dt = \frac{1}{T} \int_0^T \left| \frac{d}{dt} \left( \sqrt{x(t)^2 + \left( \frac{1}{\pi} \text{P.V.} \int_{-\infty}^{\infty} \frac{x(\tau)}{t-\tau} d\tau \right)^2} \right) \right| dt$$

To incorporate self-organized criticality (SOC), power laws, and chaos theory, we extend this to consider the statistical distributions and scaling behaviors:

$$\langle \omega \rangle = \frac{\int_0^\infty \omega \left| \mathcal{F}\{\phi(t)\}(\omega) \right|^2 d\omega + \int_{\mathbb{R}^3} \mathbf{k} \cdot \nabla \left| \psi(\mathbf{r}, t) \right|^2 d^3 r}{\int_0^\infty \left| \mathcal{F}\{\phi(t)\}(\omega) \right|^2 d\omega + \int_{\mathbb{R}^3} \left| \psi(\mathbf{r}, t) \right|^2 d^3 r}$$

Where:

- $R_e$ is the evolution rate, representing the mean absolute rate of change of the envelope.
- $T$ is the total time period of the signal.
- $A(t)$ is the envelope of the signal $x(t)$.
- $x(t)$ is the time-domain signal.
- $\mathcal{H}\{x(t)\}$ is the Hilbert transform of $x(t)$, providing the imaginary part of the analytic signal.
- $\langle \omega \rangle$ is the spectral centroid, representing the average frequency of the system.
- $\omega$ is the angular frequency.
- $\phi(t)$ is the time-domain signal representing the system's state.
- $\mathcal{F}\{\phi(t)\}(\omega)$ is the Fourier transform of $\phi(t)$, mapping the time-domain signal to the frequency domain.
- $\psi(\mathbf{r}, t)$ is the wave function representing the system's state in both space and time.
- $\mathbf{k}$ is the wave vector.
- $\mathbf{r}$ is the position vector in three-dimensional space.
- P.V. denotes the Cauchy principal value, ensuring proper handling of singularities in the Hilbert transform.

*Equation 9: Integrating Self-Organized Criticality and Chaotic Theory in Signal Evolution Analysis*

As further understanding of SOC in neural systems is developed, new possibilities for designing brain-inspired computing architectures that can harness the power of criticality emerge. These advancements could lead to the development of more efficient, adaptive, and robust artificial intelligence systems that more closely mimic the information-processing capabilities of biological systems.

**Application in Game Actions**

The BSIS software includes a component for translating complex neural signal features into actionable commands for applications such as interactive games. This functionality is encapsulated in the FeaturesToGameAction class, which maps extracted neural features to specific game actions. The FeaturesToGameAction class allows for dynamical adjustments to the game actions based on real-time changes in neural signal features, providing a responsive and adaptive control mechanism. The FeaturesToGameAction class listens to the "EXTRACTED FEATURES" topic for incoming neural signal features and publishes game actions to the "GAME ACTIONS" topic using

MQTT [23]for communication. A GUI built with PyQt5 [21] displays the neural signal features, their values, and the corresponding actions taken, updating dynamically based on incoming data. The class processes a core set of neural signal features, including Higuchi fractal dimension, evolution rate, spectral centroids, spectral edge density, variance, standard deviation, root mean square (RMS), and peak count. Instead of using predefined mappings for actions, the script compares the current feature values to previously recorded ones. This approach accounts for individual variability in neural signals and can adapt to changes in a user's neural patterns over time. The class continuously updates its internal state based on incoming neural data, ensuring that the game actions reflect the most current neural activity.

**Dynamic Adjustment of Force in Interactive Systems**

The dynamic adjustment of force based on changes in neural signal features can be deeply understood through the lens of continuous signal processing and control theory. These systems process information in a manner that mimics biological neural networks, allowing for a more natural and fluid integration of neural data into interactive applications. Continuous signal processing enables the capture of subtle variations in neural activity, providing a richer and more nuanced understanding of the underlying dynamics. The theoretical foundation for this approach lies in the study of continuous dynamical systems, capturing the inherent complexity and adaptability of neural processes. By leveraging these principles, systems can dynamically adjust interactive forces in response to real-time neural signals, creating a seamless and responsive user experience.

The feedback mechanism is crucial for adjusting forces based on neural signal changes. Feedback control theory, well-established in both engineering and neuroscience, provides the mathematical framework for this dynamic adjustment. Feedback loops continuously monitor the neural signal features and compare them to previous values. Any deviation from these values triggers an adjustment in the force applied in the interactive environment. This feedback mechanism is analogous to biological homeostasis, where organisms maintain internal stability through continuous monitoring and adjustment of physiological parameters. This continuous feedback ensures that the system can respond smoothly and in real-time to the varying states of neural activity.

The integration of neural signal features in the BSIS involves understanding the biophysical properties of neural activity. Neural signals are electrical in nature, generated by the ion flow across cell membranes. The Hodgkin-Huxley model describes this process mathematically:

$$C\frac{dV}{dt} = -g_{\text{Na}}(V - E_{\text{Na}}) - g_{\text{K}}(V - E_{\text{K}}) - g_{\text{L}}(V - E_{\text{L}}) + I$$

Where:
- $C$ is the membrane capacitance.
- $V$ is the membrane potential
- $g_{\text{Na}}, g_{\text{K}}, g_{\text{L}}$ are the conductances for sodium (Na), potassium (K), and leakage (L) ions, respectively.
- $E_{\text{Na}}, E_{\text{K}}, E_{\text{L}}$ are the reversal potentials for sodium (Na), potassium (K), and leakage (L) ions, respectively.
- $I$ is the external current.

*Equation 10: Hodgkin-Huxley Model for Membrane Potential Dynamics*

Changes in neural signal features, such as variance and RMS, reflect alterations in these ionic currents and membrane potentials. Continuous processing capabilities enables real-time monitoring and interpretation of biophysical changes in neural signals, enabling precise and timely adjustments in interactive forces. This biophysical perspective is essential for developing systems that can effectively interface with neural data. By understanding these underlying biophysical equations, the system can map neural signal changes to appropriate force adjustments, ensuring the system's response aligns with the underlying neural dynamics. The integration of this biophysical knowledge enhances the system's ability to adapt to the complex and dynamic nature of neural activity.

Control systems theory enhances the adaptability and responsiveness of interactive applications. Control systems operate on continuous feedback, allowing for smooth and precise adjustments in response to neural signal changes. This continuous feedback loop is essential for maintaining the dynamic balance required in interactive environments. By employing control theory, systems can modulate interactive forces based on real-time neural data, ensuring that the system remains responsive and adaptive to the user's neural state. This approach not only enhances the user experience but also provides insights into the dynamic nature of neural activity.

Incorporating the concept of self-organized criticality (SOC) into these systems further deepens our understanding of neural dynamics. SOC describes how complex systems naturally evolve to a critical state where small perturbations can lead to significant changes. Neural networks often exhibit SOC, where spontaneous activity can scale across different levels of organization. With continuous processing capabilities, systems are well-suited to capture these scale-invariant properties. By monitoring changes in neural signal features, these systems can dynamically adjust forces to reflect the criticality and complexity of neural activity, providing a realistic and immersive interactive experience.

The practical implementation involves real-time processing and continuous adaptation. Neural signals are continuously monitored, and their features are analyzed to detect changes in real time. The system then adjusts the force applied in the interactive environment accordingly. This continuous adaptation ensures that the system remains responsive to the user's neural state, creating a seamless interaction between the neural data and the external application. This real-time processing capability highlights the potential for developing advanced interactive systems that can adapt to the user's neural dynamics.

The interactive force dynamics with neural signal evolution rate can be described as:

$$\mathbf{F}(t) = \mathbf{K}\left(\mathbf{x}(t) - \mathbf{x}_d(t)\right) + \mathbf{B}\left(\frac{d\mathbf{x}(t)}{dt} - \frac{d\mathbf{x}_d(t)}{dt}\right)$$
$$+ \mathbf{M}\left(\frac{d^2\mathbf{x}(t)}{dt^2} - \frac{d^2\mathbf{x}_d(t)}{dt^2}\right)$$
$$+ \alpha \int_0^t \left(\frac{d}{dt}\left(\sqrt{x(t)^2 + \left(\frac{1}{\pi}\text{P.V.}\int_{-\infty}^\infty \frac{x(\tau)}{t-\tau}d\tau\right)^2}\right)\right) dt$$

(1)

Where:
- $\mathbf{F}(t)$ is the interactive force applied at time $t$.
- $\mathbf{x}(t)$ is the current state vector of the system.
- $\mathbf{x}_d(t)$ is the desired state vector of the system.
- $\mathbf{K}$, $\mathbf{B}$, and $\mathbf{M}$ are the stiffness, damping, and mass matrices, respectively.
- $\alpha$ is a scaling factor for the evolution rate.
- $x(t)$ is the time-domain neural signal.
- $\mathcal{H}\{x(t)\}$ is the Hilbert transform of $x(t)$.
- $\frac{1}{\pi}\text{P.V.}\int_{-\infty}^\infty \frac{x(\tau)}{t-\tau}d\tau$ is the Cauchy principal value of the Hilbert transform, ensuring proper handling of singularities.
- $\frac{d}{dt}\left(\sqrt{x(t)^2 + \mathcal{H}\{x(t)\}^2}\right)$ is the rate of change of the envelope of the signal.

*Equation 11: Interactive Force Dynamics with Neural Signal Evolution Rate*

The dynamic adjustment of force based on neural signal features is rooted in deep theoretical principles from continuous dynamical systems, control theory, and neurophysics. By leveraging continuous processing capabilities, these systems can accurately and responsively interface with real-time neural data. This integration not only enhances interactive applications but also provides profound insights into the dynamic and complex nature of neural activity. The application in this context represents a significant advancement in our ability to create adaptive and immersive interactive environments that respond to the user's neural state in real time. This approach bridges the gap between theoretical neuroscience and practical applications, demonstrating the profound potential in advancing our understanding and utilization of neural signals.

**Analogue AI and Our System**

Our system's functionality mirrors analogue AI in its continuous signal processing and adaptive response mechanisms. In analogue AI, information is processed as continuous waveforms, capturing the natural, uninterrupted flow of data. Similarly, our system processes neural signals in their continuous form, which allows for the detection of intricate variations in neural activity that would be lost in a discrete, digital approach. This continuous processing is crucial for accurately reflecting the dynamic and complex nature of neural signals, which are inherently variable and rich in detail. By maintaining the continuity of the data, our system can provide a more detailed and precise analysis, akin to how biological neural networks operate, capturing the subtle nuances of neural activity in real time.

Further paralleling analogue AI, our system utilizes continuous feedback loops to dynamically adjust interactive forces based on real-time neural signals. This feedback mechanism ensures that the system can adapt to changes in neural activity with high precision, maintaining a responsive and immersive user experience. In analogue AI, such feedback loops are essential for achieving a natural and adaptive interaction with the environment. Similarly, our system's feedback mechanism continuously monitors neural signal features, compares them to previous states, and makes real-time adjustments to the interactive forces. This process is crucial for creating a seamless interaction between the neural data and the external application, mirroring the adaptive and real-time processing capabilities of analogue AI. By leveraging these continuous feedback loops, our system can respond to the dynamic nature of neural activity, providing an experience that is both responsive and deeply integrated with the user's neural state.

The integrative force model with neural, synaptic, and quantum dynamics can be described as:

$$\begin{aligned}
\mathbf{F}(t) = {} & \mathbf{K}\left(\mathbf{x}(t) - \mathbf{x}_d(t)\right) + \mathbf{B}\left(\frac{d\mathbf{x}(t)}{dt} - \frac{d\mathbf{x}_d(t)}{dt}\right) \\
& + \mathbf{M}\left(\frac{d^2\mathbf{x}(t)}{dt^2} - \frac{d^2\mathbf{x}_d(t)}{dt^2}\right) \\
& + \gamma \int_0^t \left(\sqrt{\left(\frac{dx(t)}{dt}\right)^2 + \left(\frac{d\mathcal{H}\{x(t)\}}{dt}\right)^2}\right) dt \\
& + \delta \int_0^t \left(\mathcal{L}\{x(t), \mathcal{F}\{x(t)\}, \phi(t), \mathbf{V}, \mathbf{W}, H(\phi(t)), \Lambda(t)\}\right) dt \\
& + \eta \sum_{n=1}^N \frac{1}{t_n} \sum_{m=1}^M \frac{1}{(t_m - t_n)^2 + \epsilon^2} \\
& + \zeta \int_0^t \left(\frac{d}{dt}\left(\sigma(\mathbf{V}, \mathbf{W}, t) \cdot \mathcal{A}(\mathbf{V}, \mathbf{W}, t)\right)\right) dt \\
& + \theta \sum_{i=1}^N \left(\frac{\partial^2 \mathbf{\Psi}_i}{\partial t^2} + \alpha_i \frac{\partial \mathbf{\Psi}_i}{\partial t} + \beta_i \mathbf{\Psi}_i\right) \\
& + \lambda \int_0^t \left(\mathcal{Q}\{x(t), \mathcal{H}\{x(t)\}, \phi(t)\}\right) dt \\
& + \mu \int_0^t \left(\mathcal{F}^{-1}\{\xi(\omega, t)\}\right) dt \\
& + \nu \int_0^t \left(\mathbf{D}(\mathbf{r}, t) \cdot \nabla \mathbf{U}(\mathbf{r}, t)\right) dt
\end{aligned}$$

*Equation 12: Integrative Force Model with Neural, Synaptic, and Quantum Dynamics*

events with a small parameter $\epsilon$ ensuring stability, reflecting SOC.

- $t_n$ and $t_m$ are the discrete times of observed neural events, reflecting SOC.
- $\sigma(\mathbf{V}, \mathbf{W}, t)$ is the synaptic weight function, incorporating both analogue and digital synaptic adjustments.
- $\mathcal{A}(\mathbf{V}, \mathbf{W}, t)$ is the activation function of neurons, influenced by synaptic weights and neural activity.
- $\mathbf{V}$ and $\mathbf{W}$ are the vectors of synaptic inputs and weights, respectively.
- $\sum_{i=1}^{N} \left( \frac{\partial^2 \mathbf{\Psi}_i}{\partial t^2} + \alpha_i \frac{\partial \mathbf{\Psi}_i}{\partial t} + \beta_i \mathbf{\Psi}_i \right)$ captures the second-order differential equation modeling the dynamic response of each neural signal component $\mathbf{\Psi}_i$ with damping $\alpha_i$ and stiffness $\beta_i$.
- $\mathcal{Q}\{x(t), \mathcal{H}\{x(t)\}, \phi(t)\}$ represents a quantum-inspired transformation function incorporating the neural signal, its Hilbert transform, and the neural potential.
- $\mathcal{F}^{-1}\{\xi(\omega, t)\}$ is the inverse Fourier transform of a frequency-domain function $\xi(\omega, t)$ representing spectral edge density.
- $\mathbf{D}(\mathbf{r}, t) \cdot \nabla \mathbf{U}(\mathbf{r}, t)$ represents a term from fluid dynamics, capturing multi-scale interactions with the displacement field $\mathbf{U}(\mathbf{r}, t)$.

- $\mathbf{x}_d(t)$ is the desired state vector of the system.
- $\mathbf{K}$, $\mathbf{B}$, and $\mathbf{M}$ are the stiffness, damping, and mass matrices, respectively.
- $\gamma$, $\delta$, $\eta$, $\zeta$, $\theta$, $\lambda$, $\mu$, and $\nu$ are scaling factors for the different components of the system.
- $x(t)$ is the time-domain neural signal.
- $\frac{dx(t)}{dt}$ and $\frac{d\mathcal{H}\{x(t)\}}{dt}$ are the time derivatives of the neural signal and its Hilbert transform, capturing the rate of change.
- $\mathcal{H}\{x(t)\}$ is the Hilbert transform of $x(t)$.
- $\mathcal{F}\{x(t)\}$ is the Fourier transform of $x(t)$, representing the signal in the frequency domain.

events with a small parameter $\epsilon$ ensuring stability, reflecting SOC.
- $t_n$ and $t_m$ are the discrete times of observed neural events, reflecting SOC.
- $\sigma(\mathbf{V}, \mathbf{W}, t)$ is the synaptic weight function, incorporating both analogue and digital synaptic adjustments.
- $\mathcal{A}(\mathbf{V}, \mathbf{W}, t)$ is the activation function of neurons, influenced by synaptic weights and neural activity.
- $\mathbf{V}$ and $\mathbf{W}$ are the vectors of synaptic inputs and weights, respectively.
- $\sum_{i=1}^{N} \left( \frac{\partial^2 \Psi_i}{\partial t^2} + \alpha_i \frac{\partial \Psi_i}{\partial t} + \beta_i \Psi_i \right)$ captures the second-order differential equation modeling the dynamic response of each neural signal component $\Psi_i$ with damping $\alpha_i$ and stiffness $\beta_i$.
- $\mathcal{Q}\{x(t), \mathcal{H}\{x(t)\}, \phi(t)\}$ represents a quantum-inspired transformation function incorporating the neural signal, its Hilbert transform, and the neural potential.
- $\mathcal{F}^{-1}\{\xi(\omega, t)\}$ is the inverse Fourier transform of a frequency-domain function $\xi(\omega, t)$ representing spectral edge density.
- $\mathbf{D}(\mathbf{r}, t) \cdot \nabla \mathbf{U}(\mathbf{r}, t)$ represents a term from fluid dynamics, capturing multi-scale interactions with the displacement field $\mathbf{U}(\mathbf{r}, t)$.

**Game Simulation: Shuffleboard Game**

An example of BSIS's application is the shuffleboard game simulation, which translates neural signal features into game actions. The shuffleboard game is a one dimensional game, consisting of pushing a shuffleboard puck in one direction, using the force of the throw to equal the distance, and the distance to the target being included in the outcome of the game. The shuffleboard puck is pushed every second, giving adequate time for the rat to consume the reward solution. The game interface provides real-time feedback on the player's performance, displaying the target and actual distances achieved by the player. The game window updates dynamically based on the neural signal analysis, allowing players to see the immediate impact of their neural activity on the game. The system also includes functionality for logging historic data, ensuring that all game sessions and neural activities are recorded for future analysis. This is achieved through an MQTT [23]client that subscribes to the "historic_data" topic, capturing detailed logs of each game session and saving them to a designated directory. The logged data includes game duration, actions taken, results achieved, and timestamps, providing a comprehensive record of each session. The BSIS integrates feedback mechanisms to reinforce desired behaviors during neural training. The FeedbackSystem class plays a crucial role in this process by providing auditory and physical rewards based on the outcome of neural activities. Positive reinforcement is delivered through a combination of reward sounds and activation of a feeder mechanism controlled by a USB relay. In contrast, distress sounds (inaudible to humans) are used as a deterrent for undesired behaviors. These feedback mechanisms ensure that the training process is effective and humane. These feedback signals are transmitted back to the brain through the MEA, completing the bidirectional communication loop. This loop ensures continuous interaction between the biological and silicon components, facilitating real-time adaptation and learning.

**Digital Stimulation Metadata Encoding**

The BSIS offers users the option to utilize digital stimulation, which involves encoding neural feedback into digital signals for precise control. The digital stimulation metadata encoder translates game state features into binary signals for transmission as ON (150 microvolts) or OFF (0 microvolts) signals. Features like score, round, distance to target, and player forces are converted into binary signals used to generate stimulation waveforms based on the outcome of neural activities, such as reward or distress. The encoding method converts each feature value into a binary string representation. Each feature value is scaled to fit within a specified range and then converted to a binary string. This binary string is translated into a sequence of ON and OFF signals, where the ON signal corresponds to 150 microvolts and the OFF signal to 0 microvolts. For example, a feature value is first normalized to a range between 0 and 100, then scaled to fit within a 16-bit binary representation. Each bit in the binary representation is converted to an ON or OFF signal based on its value (1 or 0), generating a series of digital signals that represent the encoded feature value. The generated digital signals are then combined to create a stimulation waveform. The waveform is adjusted based on the latest neural activity outcome, with distinct waveforms for different outcomes such as reward or distress. A reward outcome generates a clean sine wave, while a distress outcome generates a chaotic wave. The final waveform is composed by appending the outcome-based waveform to the feature-encoded signals. This method of encoding and transmitting neural feedback as digital signals enables BSIS to effectively control external devices and applications, providing a robust interface for real-time neural signal integration.

**Digital Stimulation Signal Output**

The digital stimulation signal outputter is responsible for taking encoded digital signals and transmitting them to stimulation hardware. This process ensures the signals are correctly formatted and continuously sent, enabling the neural feedback system to operate effectively. To transmit the waveforms, the outputter uses PyAudio, a library for audio playback. The system sets up audio streams for each channel, specifying the output device indices. The audio callback function continuously feeds the waveform data from ring buffers to the audio streams, ensuring smooth and uninterrupted playback. Each channel has a dedicated ring buffer that holds the waveform data, with positions managed cyclically to ensure seamless signal transmission. The outputter includes a visualization component built with PyQt5 and PyQtGraph[21,22], providing a real-time display of the digital waveforms being transmitted for monitoring and verification of the signal output. The core components of the digital stimulation signal outputter include MQTT communication[23], waveform conversion, continuous signal buffering, audio playback, and real-time visualization. The on_connect and on_message functions handle MQTT communication[23], subscribing to the relevant topic and processing incoming messages. The waveform conversion translates binary values to amplitude values, while the ring buffers ensure continuous playback. PyAudio manages the audio streams, and the visualization component provides real-time feedback on the waveforms. Overall, the digital stimulation signal outputter integrates these components to create a robust system for transmitting neural feedback signals to stimulation hardware. This integration ensures the neural feedback system can operate smoothly, providing precise and continuous stimulation based on the encoded digital signals.

The integrated model for generating digital stimulation signals can be described as follows:

$$\mathbf{S}(t) = \sum_{i=1}^{N} \left[ \text{Bin} \left( \frac{f_i - \min(f)}{\max(f) - \min(f)} \times (2^k - 1) \right) \times (\mathcal{O}(f_i, t) - \mathcal{X}(f_i, t)) \right]$$
$$+ \mathcal{W}(t) + (\mathbf{S}_{\text{reward}}(t) \cdot \mathcal{R}(t) + \mathbf{S}_{\text{distress}}(t) \cdot \mathcal{D}(t))$$
$$+ \int_0^t \mathcal{M}\left(\mathcal{F}_{\text{audio}}(t) + \mathcal{C}(t)\right) dt + \mathcal{V}(t)$$

Where:

- $\mathbf{S}(t)$ is the digital stimulation signal at time $t$.
- $f_i$ are the feature values such as score, round, distance to target, and player forces.
- $\text{Bin}(x)$ is the binary encoding function that converts $x$ to a binary string.
- $\mathcal{O}(f_i, t)$ is the ON signal function, representing 150 microvolts.
- $\mathcal{X}(f_i, t)$ is the OFF signal function, representing 0 microvolts.
- $\mathcal{W}(t)$ is the waveform function combining the binary signals into a stimulation waveform.
- $\mathbf{S}_{\text{reward}}(t)$ is the clean sine wave for reward outcomes.
- $\mathcal{R}(t)$ is the reward indicator function.
- $\mathbf{S}_{\text{distress}}(t)$ is the chaotic wave for distress outcomes.
- $\mathcal{D}(t)$ is the distress indicator function.
- $\mathcal{M}(x)$ is the audio modulation function.
- $\mathcal{F}_{\text{audio}}(t)$ is the audio stream function.
- $\mathcal{C}(t)$ is the continuous signal buffering function.
- $\mathcal{V}(t)$ is the visualization function providing real-time display of digital waveforms.

*Equation 13: Integrated Model for Generating Digital Stimulation Signals*

The BSIS proprietary system design is as follows:

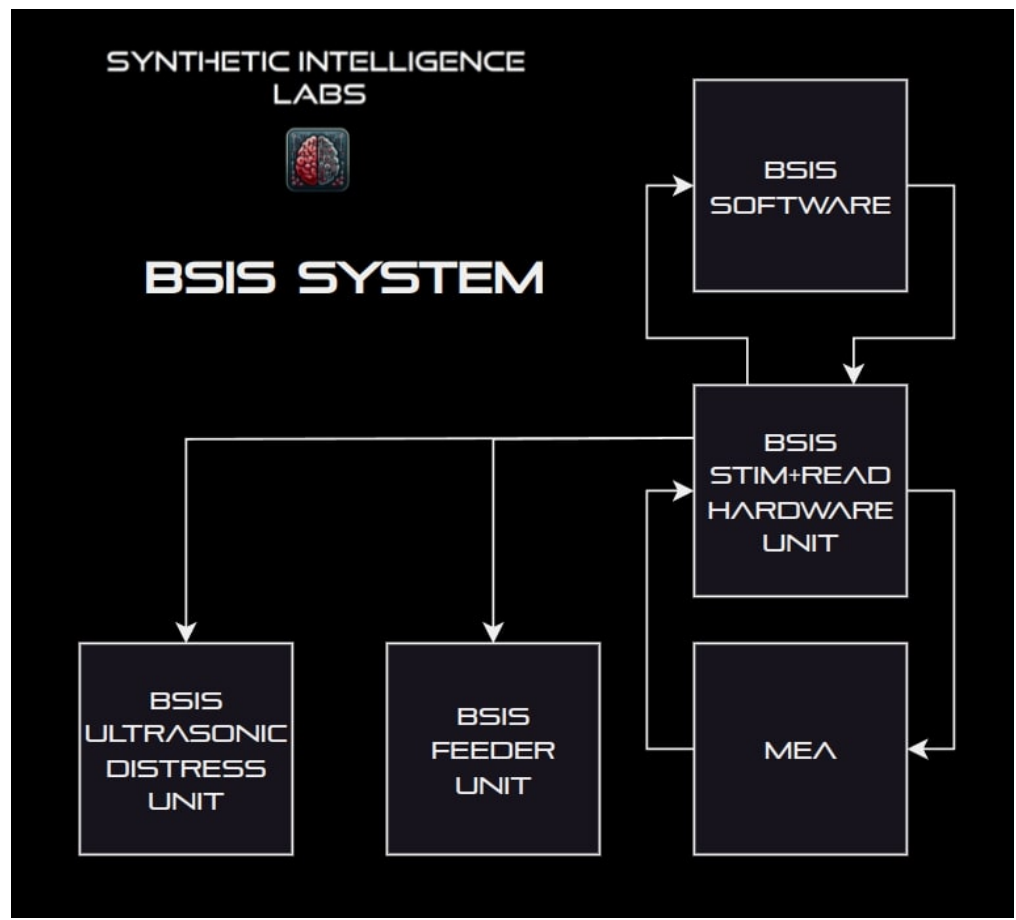

*Figure 13: BSIS Proprietary System Design*

Overall the BSIS system can be described as follows:

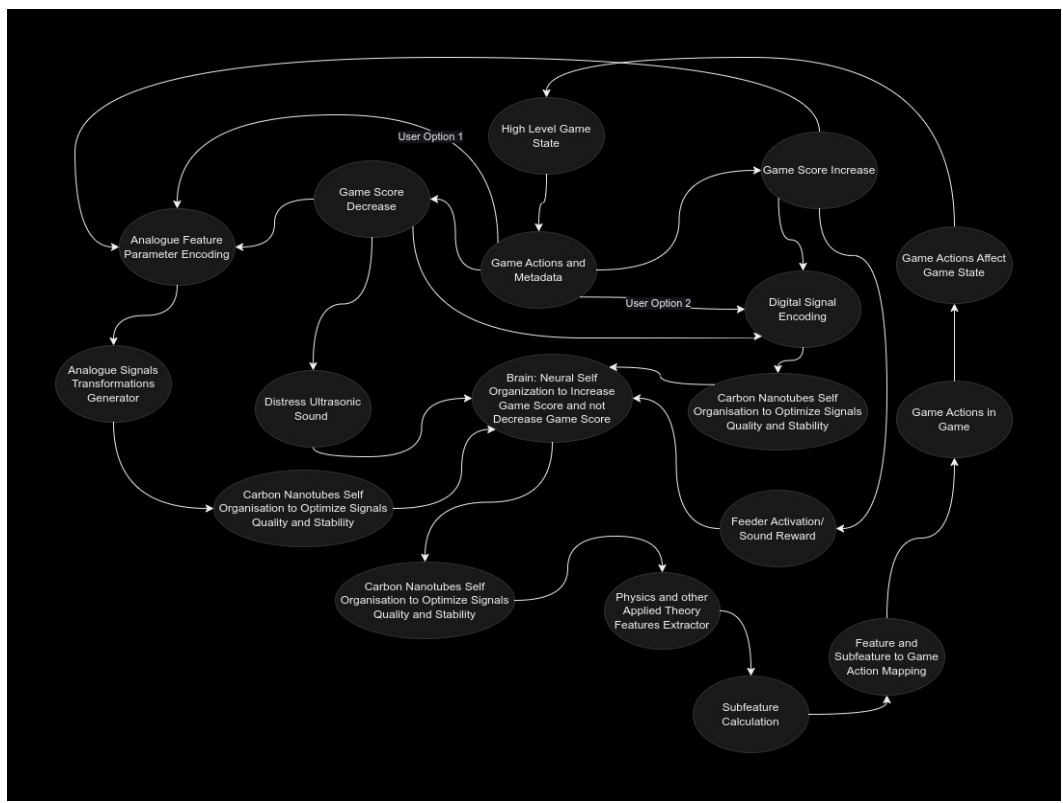

*Figure 14: Diagram of BSIS Physics Informed Hybrid Hierarchical Reinforcement Learning State Machine*

**Analogue Stimulation Metadata Encoding**

BSIS also provides an analogue option for stimulation, involving the encoding of neural feedback into analogue signals. The analogue metadata encoder translates game state features into various parameters used to generate these signals. The encoding process ensures the effective transformation of neural feedback into analogue signals. First, the system processes game state features, such as distance to target and player force, normalizing these values for consistency across different input ranges. This normalization allows uniform processing of features regardless of their original scale. The encoder then determines the system's state (e.g., resting or active) based on the normalized distance, tailoring feedback signals to the current game context. Key features like variability factor, variance, standard deviation, RMS value, fractal dimension, and peak height are encoded using specific algorithms. For example, variance is calculated as the sum of normalized distance and force, while standard deviation is a weighted average of these normalized values. Similarly, other features are derived through defined mathematical relationships involving the normalized input values. These encoded features are organized per channel, with each channel receiving a set of encoded parameters. Features like variability factor, variance, and RMS value are calculated for each channel independently, ensuring comprehensive stimulation signals that cover various aspects of neural activity. To transmit these features, the encoder uses the MQTT protocol[23], publishing encoded features at regular intervals (4 Hz). This ensures the feedback loop remains responsive and up-to-date with the latest game state. The MQTT[23] messages contain all encoded features, structured for easy decoding and use by the receiving end to generate appropriate analogue signals. Finally, the encoded features are sent as MQTT[23] messages, allowing continuous and accurate generation of analogue signals. This variation in analogue signals is particularly useful for applications requiring a more nuanced and adaptable form of neural stimulation.

**Analogue Signal Generation**

The analogue signal generator processes the encoded metadata to produce ECoG-like signals used for neural stimulation. This involves several steps to ensure the generated signals are complex and representative of real neural activity. The process begins by generating base signals within a specified voltage range, either starting from initial values or generating them randomly within the range. These base signals are then scaled to a specific bit depth, normalizing the signal values between the minimum and maximum voltages. To introduce realistic neural characteristics, various transformations are applied to the signals. Oscillations corresponding to different brain states (such as resting or active) are added by generating sinusoidal patterns within specific frequency bands (e.g., alpha, beta, gamma). Amplitude variability is introduced based on a variability factor, with noise scaled by the standard deviation of the signals added to them. Variance and standard deviation adjustments ensure the signals maintain the desired statistical properties while remaining within the voltage range. The signals are scaled to have a specified RMS value, and peaks are added at random positions, scaled appropriately to fit within the voltage range. Fractal structures are introduced using the Hurst exponent derived from the fractal dimension, creating patterns that mimic the complexity of neural signals. Zero-crossing rates are adjusted to target rates by scaling the signals, and Arnold tongues are applied by generating sinusoidal patterns with added harmonics and noise. Phase synchronization is achieved by adjusting the phase of the signals' Fourier transform components, and transfer entropy involves adding influenced signals based on interaction weights derived from the influence factor. The Hilbert-Huang transform modulates the signals with low and high-frequency components, which are then clipped to the voltage range. Spectral centroids and edge densities are applied by adjusting the power spectrum of the signals, while dynamic time warping aligns signals to themselves, ensuring proper scaling. The signals undergo FFT and are adjusted based on a complexity factor, while signal evolution involves progressively adding random variations. Phase-amplitude coupling combines low-frequency and high-frequency components, and Granger causality adjusts signals based on causal relationships. Multivariate empirical mode decomposition decomposes signals into intrinsic mode functions (IMFs), which are then combined to achieve the desired structure. Lastly, quantum-inspired transformations are applied using parameterized Hermitian matrices. These matrices are generated with specified value ranges, and their eigenvalues and eigenvectors are used to create density matrices. The signals are modified based on the interpolated density matrix values, ensuring they reflect the desired quantum-inspired properties. By applying these transformations, the analogue signal generator produces complex and realistic neural-like signals suitable for stimulation, enhancing the effectiveness of the neural feedback system.

The integrated model for analogue stimulation signals can be defined as:

$$\mathbf{S}(t) = \sum_{i=1}^{N} [\mathcal{N}(f_i) \cdot \mathcal{S}(f_i) \cdot \mathcal{P}(f_i) \cdot \mathcal{V}(f_i) \cdot \mathcal{H}(f_i) \cdot \mathcal{Z}(f_i) \cdot \mathcal{A}(f_i) \cdot \mathcal{Q}(f_i)]$$
$$+ \sum_{j=1}^{M} \left[ \int_0^t \mathcal{T}_j(\mathbf{B}(t))dt + \int_0^t \mathcal{H}_j(\mathbf{C}(t))dt \right]$$
$$+ \sum_{k=1}^{L} [\mathcal{D}_k(t) + \mathcal{W}_k(t) + \mathcal{E}_k(t) + \mathcal{R}_k(t) + \mathcal{G}_k(t) + \mathcal{M}_k(t) + \mathcal{Y}_k(t) + \mathcal{K}_k(t)]$$
$$+ \sum_{p=1}^{P} [\mathcal{F}_p(t) + \mathcal{I}_p(t) + \mathcal{L}_p(t) + \mathcal{X}_p(t) + \mathcal{C}_p(t)]$$

*Equation 14: Integrated Model for Analogue Stimulation Signals*

Where:
- $\mathbf{S}(t)$ is the analog stimulation signal at time $t$.
- $f_i$ are the feature values such as distance to target and player force.
- $\mathcal{N}(f_i)$ is the normalization function for the feature values.
- $\mathcal{S}(f_i)$ is the system state determination function.
- $\mathcal{P}(f_i)$ is the parameter encoding function (e.g., variability factor, variance).
- $\mathcal{V}(f_i)$ is the variance calculation function.
- $\mathcal{H}(f_i)$ is the fractal dimension function.
- $\mathcal{Z}(f_i)$ is the zero-crossing rate function.
- $\mathcal{A}(f_i)$ is the amplitude variability function.
- $\mathcal{Q}(f_i)$ is the quantum-inspired transformation function.
- $\mathbf{B}(t)$ represents the base signals within a specified voltage range.
- $\mathbf{C}(t)$ are the continuous signal transformations (e.g., Hilbert-Huang transform).
- $\mathcal{T}_j(\mathbf{B}(t))$ is the transfer entropy function.
- $\mathcal{H}_j(\mathbf{C}(t))$ is the Hilbert-Huang transform function.
- $\mathcal{D}_k(t)$ is the dynamic time warping function.
- $\mathcal{W}_k(t)$ is the waveform generation function (e.g., sinusoidal patterns).
- $\mathcal{E}_k(t)$ is the signal evolution function.
- $\mathcal{R}_k(t)$ is the RMS value calculation function.
- $\mathcal{G}_k(t)$ is the Granger causality function.
- $\mathcal{M}_k(t)$ is the multivariate empirical mode decomposition function.
- $\mathcal{Y}_k(t)$ is the phase synchronization function.
- $\mathcal{K}_k(t)$ is the spectral centroid and edge density adjustment function.
- $\mathcal{F}_p(t)$ is the fractal structure function.
- $\mathcal{I}_p(t)$ is the influence factor function.
- $\mathcal{L}_p(t)$ is the low-frequency modulation function.

**Detailed Transformations:**

- Scaling signals to bit depth
- Applying oscillations based on system state
- Introducing amplitude variability and variance
- Adjusting signals to match standard deviation and RMS values
- Adding peaks and applying moving averages
- Incorporating fractal structures using Hurst exponent
- Modifying signals to match zero-crossing rates
- Applying Arnold tongues, phase synchronization, and transfer entropy
- Performing Hilbert-Huang transform and adjusting spectral centroids
- Utilizing dynamic time warping, FFT, and signal evolution techniques
- Implementing phase-amplitude coupling and Granger causality
- Conducting multivariate empirical mode decomposition
- Applying quantum-inspired transformations using parameterized Hermitian matrices

**Stimulation Hardware**

The audio outputs from BSIS are then run through voltage dividers to adjust signal levels to the appropriate range for safe and effective neural stimulation in the millivolt range, before being directed to the MEA. Additionally, all shielding for wires and components enclosures are grounded to minimize electromagnetic interference and ensure stable signal transmission. This comprehensive approach to hardware design ensures high-fidelity signal acquisition, processing, and feedback, creating a robust platform for neural interfacing. The signal processing and analysis framework of BSIS ensures high-fidelity capture, precise processing, and comprehensive analysis of neural signals. Through advanced preprocessing, feature extraction, and detailed analysis, the system provides actionable insights and effective feedback, supporting advanced research in neurotechnology and bioengineering.

## 4. Discussion

**Further Development of System**

The Bio-Silicon Intelligence System (BSIS) could be significantly enhanced by integrating a multitude of advanced computational theories to manage complex neural signal processing and adaptive game control. Building on the foundation of Hierarchical

Reinforcement Learning (HRL) and Hybrid Automata, the system effectively manages both discrete and continuous dynamics. Continuous dynamics of neural signals are currently captured through a vast array of features, such as peak count, spectral centroid, and Higuchi Fractal Dimension. We are also developing graph neural networks that utilize various topological data embeddings to further enhance the processing and interpretation of complex neural signals. Adaptive Dynamic Programming could optimize control and movement strategies within the game, ensuring efficient mapping of physical actions to neural inputs. Lagrangian Mechanics could be employed to model the system's physical dynamics and optimize the force and movement parameters for game actions.

To formalize the symbolic representation and processing of neural features, BSIS could employ Symbolic Computation and Formal Languages, including Automata Theory and Petri Nets. These frameworks could enhance the clarity and precision of brain-computer interactions by structuring the symbolic encoding and decoding of neural signals. Symbolic Computation could be used to create complex expressions that encapsulate neural data patterns, while Automata Theory and Petri Nets could model the flow and processing of these symbolic data. Category Theory and Monads could provide a robust mathematical framework to handle the compositional structures of these symbolic expressions, facilitating more complex and flexible data transformations. Fuzzy Logic and Probabilistic Graphical Models could be used to manage the inherent uncertainty and variability in neural data, allowing the system to make informed decisions even with noisy inputs.

Graph Theory and Information Geometry could be employed to map and analyze the intricate relationships between neural features, game actions, and environmental metadata, providing deeper insights into the underlying structures of neural data. This could help in visualizing and optimizing the pathways through which neural signals are processed and translated into game actions. Reinforcement Learning with Function Approximation could enhance the system's learning capabilities, allowing it to generalize from limited data and perform complex computations efficiently. Process Calculus could provide a formal framework for modeling concurrent processes and interactions within the system. Denotational and Operational Semantics, along with Automated Theorem Proving, could ensure the correctness and reliability of symbolic transformations and state transitions within the system.

The development of a self-optimizing language for brain-computer and computer-brain communication could be central to BSIS. Symbolic Communication and Computational Linguistics, along with Formal Grammars such as Context-Free Grammars, Attribute Grammars, and Tree Adjoining Grammars, could structure the symbolic encoding and decoding of neural features. These grammars could define the syntax and semantics for the symbolic language, ensuring that neural data is accurately and meaningfully translated into game actions. Phase space analysis could be utilized to understand the dynamic behavior of neural signals over time, providing a deeper understanding of the system's state transitions and stability. Hierarchical Temporal

Memory could be used to capture temporal patterns in neural data, while Deep Learning techniques such as Autoencoders could extract and compress relevant features, enhancing the system's ability to identify and utilize critical information. Symbolic Regression could be employed to discover mathematical relationships within the neural data, while automated Formal Concept Analysis could refine these symbolic representations, enabling more precise and effective communication.

Self-Organizing Maps and Neural Turing Machines could facilitate the self-organization and continuous learning of the system, ensuring it adapts to new data and scenarios effectively. Evolutionary Algorithms and Genetic Programming could be used to evolve symbolic expressions that represent neural features and their transformations, supporting the development of a robust and adaptive symbolic language. These methods could help the system to continuously improve and optimize its symbolic language, making it more efficient and expressive over time. Meta-Learning could enable the system to optimize its learning process, continuously improving its performance. Active Inference and Predictive Coding could enhance the system's ability to predict and adapt to changes in neural signals, ensuring real-time responsiveness.

Chaos Theory and Fuzzy Systems could manage the inherent complexity and variability of neural data, providing robust frameworks for signal processing and state transitions. Information-Theoretic Optimization and mutual information could optimize the encoding and transmission of neural data, ensuring that critical information is preserved and effectively communicated. Formal Verification and Validation, along with Semantic Web Technologies and Pi-Calculus, could ensure the reliability and scalability of the system, enabling it to handle complex and large-scale neural data efficiently. Temporal Logic and Formal Verification techniques could provide the tools to validate the system's behavior over time, ensuring that it operates correctly under all conditions.

By incorporating these advanced methods, the BSIS could achieve sophisticated symbolic communication, self-organization, and optimization, significantly enhancing its functionality and potential applications in AI and neurotechnology. The integration of Behavioral Trees and the use of Lagrangian Mechanics could further refine the system's decision-making and physical modeling capabilities. Graph Theory and Information Geometry could provide deep insights into the relationships between different neural features and game actions.

**Applications and Use Cases for AI**

BSIS addresses critical limitations in current AI hardware and the physical constraints associated with the miniaturization of silicon-based chips and digital AI limitations, leveraging the superior learning capabilities of biological intelligence. As traditional AI systems approach their limits in processing power and efficiency due to the physical boundaries of chip size and thermal management, BSIS offers a hybrid approach that integrates biological neural networks with silicon-based computing. This integration harnesses the adaptive and learning capabilities of biological systems, paving the way for the development of more efficient, scalable, and intelligent AI solutions.

Potential applications for BSIS span a wide range of sectors, including autonomous systems and robotics, where the adaptive and self-organizing capabilities of BSIS could lead to more robust and efficient control systems capable of operating in dynamic and unpredictable environments. In industrial automation, BSIS could optimize manufacturing processes, enhance predictive maintenance, and improve operational efficiency. Government applications could benefit from advanced data analysis, enhanced security systems, and improved decision-making capabilities. For consumers, BSIS could enable smarter home automation systems, more intuitive personal assistants, and enhanced user experiences in gaming and entertainment. By overcoming the limitations of current AI hardware, BSIS has the potential to revolutionize various fields by providing a more powerful and flexible framework for AI development.

Moreover, the hybrid bio-silicon approach of BSIS could significantly improve energy efficiency and reduce the ecological impact of AI systems. Biological systems are inherently more energy-efficient than traditional silicon-based processors, as they operate effectively at much lower power levels. This could lead to substantial reductions in the energy consumption of AI systems, addressing one of the major challenges of modern AI hardware. Additionally, the use of bio-silicon hybrids could reduce the need for complex and energy-intensive cooling solutions, further enhancing the sustainability of AI technologies. By integrating biological components, BSIS not only enhances computational capabilities but also contributes to the development of greener and more sustainable AI solutions.

**Applications and Use Cases in Neurotechnology Research**

The BSIS demonstrates significant potential in neurotechnology research, showcasing its versatility and capability to advance the field. One primary application is in the development and optimization of neural interfaces. The high-fidelity signal acquisition and processing capabilities of BSIS enable precise neural interfacing, facilitating detailed studies of neural activity and communication. By utilizing carbon nanotube-coated electrodes, BSIS enhances signal clarity and stability, paving the way for advancements in BCIs. This advancement allows for more accurate and efficient communication between the brain and external devices. BSIS is also instrumental in exploring cognitive enhancement techniques through neural stimulation and feedback mechanisms. The system's ability to deliver tailored neuromimetic signals shows promising results in modulating neural activity to enhance cognitive functions. BSIS also plays a crucial role in studying neural plasticity and its implications for rehabilitation. The system's dual signaling approach, combining reward solutions and distress sounds, effectively promotes neural plasticity and facilitates the reorganization of neural pathways. This is important for rehabilitation therapies aimed at recovering neural function after injury or disease, enabling targeted interventions that enhance the brain's ability to adapt and recover.

**Applications and Use Cases in Bioengineering Advancements**

BSIS has shown considerable potential in advancing bioengineering by integrating biological neural networks with silicon-based computing, opening new possibilities for research and applications. It significantly contributes to synthetic biology and tissue engineering by providing a robust platform for studying interactions between biological tissues and engineered systems. The integration of BSIS into bio-hybrid systems opens possibilities to novel interfaces that combine living tissues with artificial components. By leveraging BSIS's bidirectional communication capabilities, researchers can develop bio-hybrid devices that can perform complex functions by mimicking natural biological processes. The application of BSIS in personalized medicine has the potential to revolutionize patient care by tailoring treatments based on individual neural responses. By analyzing the neural activity of patients in response to different therapies, BSIS helps identify the most effective treatment strategies for each individual. This personalized approach ensures that patients receive therapies best suited to their specific conditions, improving treatment outcomes and reducing the risk of adverse effects. BSIS promotes ethical and sustainable research practices by minimizing the need for animal testing and ensuring the humane treatment of research animals. Its design includes protocols for the ethical treatment of animals, such as positive reinforcement and mild deterrence techniques. Combining high-fidelity neural interfacing with robust data analysis and ethical research practices, BSIS paves the way for innovative solutions and groundbreaking discoveries in bioengineering.

## 5. Conclusions

The integration of biological neural networks with silicon-based computing in BSIS offers numerous advantages, bridging the gap between the adaptability of biological systems and the computational power of silicon technologies. This hybrid approach leverages the strengths of both paradigms, enabling precise and dynamic interactions not possible with either system alone. Biological neural networks provide unparalleled flexibility and learning capabilities, adapting to new information and stimuli in ways that traditional silicon-based systems cannot. By interfacing these networks with advanced silicon computing, BSIS enhances the fidelity and complexity of data processing, enabling sophisticated real-time analysis and feedback mechanisms. This synergy opens new avenues for research and application, from advanced neuroprosthetics and cognitive enhancement to more effective neuromodulation therapies. Compared to existing systems, BSIS stands out in terms of performance, versatility, and innovation. Traditional brain-computer interfaces (BCIs) and neural recording systems often lack the bidirectional communication capabilities and high-fidelity signal processing that BSIS provides. The use of carbon nanotube-coated electrodes in BSIS ensures superior signal clarity and stability, enhancing the overall performance of the system. Moreover, the dual signaling approach—using reward solutions and human-inaudible distress sounds—offers a novel method for neural conditioning, promoting more effective learning and adaptation. In terms of versatility,

BSIS supports a wide range of applications, from bio-hybrid devices and drug testing to personalized medicine and ethical research practices. This adaptability makes it a valuable tool across multiple fields, setting it apart from more specialized or limited systems.The Bio-Silicon Intelligence System represents a significant advancement in integrating biological and silicon computing. The system's innovative approach and versatile applications hold great promise for the future of neurotechnology and bioengineering.. Continued research and development will enhance its capabilities, broadening its impact and paving the way for new scientific and medical breakthroughs.

**Supplementary Materials:** The following supporting information can be downloaded at: https://github.com/Unlimited-Research-Cooperative/Bio-Silicon-Synergetic-Intelligence-System.

**Author Contributions:** For research articles with several authors, a short paragraph specifying their individual contributions must be provided. The following statements should be used d"Conceptualization, Vincent Jorgsson, Mustaf Ahmed, Maxx Yung, Aryaman Pattnayak, Sri Pradhyumna Sridhar, Vaishnav Varma, Arun Ram Ponnambalam, Georg Weidlich, Dimitris Pinotsis; methodology, Vincent Jorgsson, Raghav Kumar.; software, Vincent Jorgsson, Raghav Kumar, Mustaf Ahmed; validation, Vincent Jorgsson, Raghav Kumar;  formal analysis, Vincent Jorgsson; investigation, Vincent Jorgsson; resources, Vincent Jorgsson; data curation, Vincent Jorgsson; writing—original draft preparation, Vincent Jorgsson; writing—review and editing, Vincent Jorgsson, Raghav Kumar, Mustaf Ahmed, Maxx Yung, Aryaman Pattnayak, Sri Pradhyumna Sridhar, Vaishnav Varma, Arun Ram Ponnambalam, Georg Weidlich, Dimitris Pinotsis; visualization, Vincent Jorgsson, Raghav Kumar; supervision, Vincent Jorgsson, Georg Weidlich, Dimitris Pinotsis; project administration, Vincent Jorgsson; All authors have read and agreed to the published version of the manuscript.

**Data Availability Statement:** The raw data supporting the conclusions of this article will be made available by the authors on request.

**Acknowledgments:** We greatly thank all who have supported our research at Synthetic Intelligence Labs.

**Conflicts of Interest:** The authors declare no conflicts of interest.

**Appendix A**

Experimental Code: SIL-BSIS-Phase1-2024

Funding and Resource Allocation ID: BSIS-RD124

Compliance and Regulatory Codes Overview: SIL-BSIS-RD2024, SIL-BSIS-OD2024, SIL-BSIS-DUC2024, SIL-BSIS-AWC2024, SIL-BSIS-ETH-MBT2024, SIL-BSIS-MBT-SEC2024, SIL-BSIS-DURC-MBT2024, SIL-BSIS-BSBS-MBT2024, SIL-BSIS-HDM-ETH2024, SIL-BSIS-DMP-PRIV2024, SIL-BSIS-DMP-CYSEC2024, SIL-BSIS-RM-ETH2024, SIL-BSIS-RM-TECH2024, SIL-BSIS-RM-PUBPER2024

This project receives direct research and development funding from Synthetic Intelligence Labs, with a designated budget code for the fiscal year 2024.This allocation supports the project's operational and research activities.

Compliance and Operational Disclosure: The findings and statements within this document solely reflect the objectives of the research initiative and do not implyendorsement by Synthetic Intelligence Labs nor its affiliates. This endeavor aims to push the boundaries of technical development and knowledge sharing. License and Property Code: The "Bio-Silicon Intelligence System" model, a proprietary development by Synthetic Intelligence Labs, is protected under the CreativeCommons Attribution-ShareAlike 4.0 International License (CC BY-SA 4.0), facilitating the legal use and modification of the work with appropriate attribution. The foundational work is stored in the Synthetic Intelligence Labs' repository. Additional permissions may be sought directly from Synthetic Intelligence Labs. Research activities are be carried out in Synthetic Intelligence Labs' facilities, equipped with advanced instrumentation for precise, reliable experimentation. The project utilizes controlled pharmaceuticals, procured in compliance with legal standards and ethical practices concerning animal research. The methodology adheres to international standards for animal research ethics, including guidelines by the Institutional Animal Care and Use Committee (IACUC) and the National Institutes of Health (NIH), ensuring ethical treatment and care of laboratory animals. The study involves continuous veterinary oversight and adherence to acomprehensive animal welfare protocol, ensuring the health and well-being of rodent subjects through systematic health assessments and maintenance of appropriate living conditions.